\documentclass[journal]{IEEEtran}

\usepackage{amsfonts}
\usepackage{amsmath}
\usepackage{amssymb}
\usepackage{graphicx}
\usepackage{mathtools}
\usepackage{adjustbox}
\usepackage{hyperref}
\hypersetup{colorlinks,allcolors=black}
\usepackage{cite} %  if labels are of an ordered list, reduce with a hyphen

\usepackage[caption=false, font=footnotesize]{subfig} % for subfigures

\usepackage{amsthm} % for the proof

\usepackage{soul}  % for highlighting a text

\usepackage[normalem]{ulem}
\newtheorem{theorem}{Theorem}
\newtheorem{lemma}{Lemma}

\usepackage{hyperref}

\usepackage[makeroom]{cancel} % To cross something

\usepackage{xcolor} % text color

\usepackage{tikz}
\usetikzlibrary{shapes.multipart, shapes.geometric, arrows.meta, positioning, fit}

\tikzstyle{block} = [rectangle, minimum width=1.5cm, minimum height=1cm, text centered, draw=black, fill=white, align=center]
\tikzstyle{arrow} = [thick, ->, >=stealth]

\tikzstyle{dashedbox} = [
  draw=black, % Border color
  line width=1pt, % Border thickness
  rectangle, % Shape
  dashed, % Border style
  inner sep=0.1cm, % Inner padding
  fill=lightgray, % Fill color
  fill opacity=0.2 % Fill opacity
]

\begin{document}
% \huge
\title{PAC Codes with Bounded-Complexity Sequential Decoding: Pareto Distribution and Code Design}
\author{Mohsen~Moradi\textsuperscript{\href{https://orcid.org/0000-0001-7026-0682}{\includegraphics[scale=0.06]{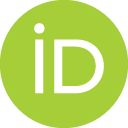}}},
Hessam~Mahdavifar\textsuperscript{\href{https://orcid.org/0000-0001-9021-1992}{ \includegraphics[scale=0.06]{figs/ORCID}}}
\thanks{The authors are with the Department of Electrical \& Computer Engineering, Northeastern University, Boston MA-02115, USA (e-mail: m.moradi@northeastern.edu, h.mahdavifar@northeastern.edu).}%
\thanks{This work was supported by NSF under Grant CCF-2415440 and the Center for Ubiquitous Connectivity (CUbiC) under the JUMP 2.0 program.}%
}

\maketitle
\begin{abstract} 
Recently, a novel variation of polar codes known as polarization-adjusted convolutional (PAC) codes has been introduced by Ar{\i}kan. These codes significantly outperform conventional polar and convolutional codes, particularly for short codeword lengths, and are shown to operate very close to the optimal bounds. It has also been shown that if the rate profile of PAC codes does not adhere to certain polarized cutoff rate constraints, the computation complexity for their sequential decoding grows exponentially.
In this paper, we address the converse problem, demonstrating that if the rate profile of a PAC code follows the polarized cutoff rate constraints, the required computations for its sequential decoding can be bounded with a distribution that follows a Pareto distribution. This serves as a guideline for the rate-profile design of PAC codes. For a high-rate PAC\,$(1024,899)$ code, simulation results show that the PAC code with Fano decoder, when constructed based on the polarized cutoff rate constraints, achieves a coding gain of more than $0.75$~dB at a frame error rate (FER) of $10^{-5}$ compared to the state-of-the-art 5G polar and LDPC codes.

\end{abstract}

%##########################################################################################################

\section{Introduction}

\IEEEPARstart{D}{elivering} short data packets with ultra-low latency and high reliability is one of the primary goals of the next-generation wireless communication systems. 
The error-correction performance of the recently introduced polarization-adjusted convolutional (PAC) codes for some short block lengths and different code rates can approach the non-asymptotic channel coding bound approximation known as the dispersion approximation \cite{arikan2019sequential, moradi2020PAC}.
This promising performance is attained with a Fano sequential decoder that has a varying complexity which could potentially become exponential in the block length. However, it has been shown that at high signal-to-noise ratios (SNRs), PAC codes can be decoded using sequential decoding with a relatively small complexity per decoded bit \cite{moradi2020PAC}.
Exploring PAC codes is an ongoing research endeavor, and designing them for longer code lengths and higher code rates while maintaining a low decoding complexity, i.e., a bounded complexity per decoded bit, is desirable.

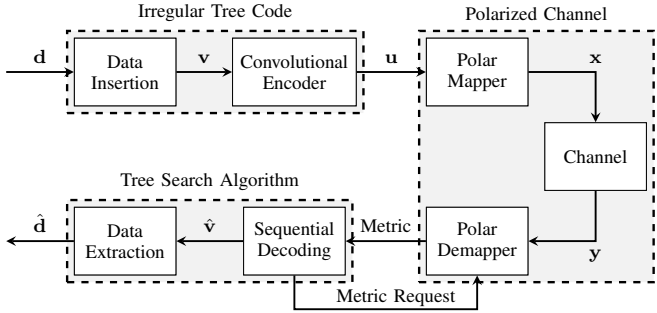
\begin{figure}[t]
\centering
\resizebox{\columnwidth}{!}{
\begin{tikzpicture}[node distance=1.5cm, every node/.append style={font=\footnotesize}]
% Nodes
\node (Insertion) [block] {Data \\ Insertion};
\node (conv_enc) [block, right of=Insertion, xshift=1cm] {Convolutional \\ Encoder};
\node (p_enc) [block, right of=conv_enc, xshift=1.2cm] {};
\node (ch) [block, right of=p_enc, xshift=.25cm, yshift=-1.25cm] {};
\node (p_dec) [block, below of=p_enc, yshift=-1cm] {};
\node (conv_dec) [block, below of=conv_enc, yshift=-1cm] {};
\node (Extraction) [block, below of=Insertion, yshift=-1cm] {};

% Larger dashed boxes around the nodes
\node[dashedbox, fit=(Insertion) (conv_enc), label={[label distance=.01cm]above:Irregular Tree Code}] {};
\node[dashedbox, fit=(p_enc) (ch) (p_dec), label={[label distance=.01cm]above:Polarized Channel}] {};
\node[dashedbox, fit=(conv_dec) (Extraction), label={[label distance=.01cm]above:Tree Search Algorithm}] {};

\node (Insertion) [block] {Data \\ Insertion};
\node (conv_enc) [block, right of=Insertion, xshift=1cm] {Convolutional \\ Encoder};
\node (p_enc) [block, right of=conv_enc, xshift=1.2cm] {Polar \\ Mapper};
\node (ch) [block, right of=p_enc, xshift=.25cm, yshift=-1.25cm] {Channel};
\node (p_dec) [block, below of=p_enc, yshift=-1cm] {Polar \\ Demapper};
\node (conv_dec) [block, below of=conv_enc, yshift=-1cm] {Sequential \\ Decoding};
\node (Extraction) [block, below of=Insertion, yshift=-1cm] {Data \\ Extraction};

% Arrows
\draw [arrow] ([xshift=-1cm]Insertion.west) -- (Insertion.west) node[midway, above] {$\mathbf{d}$}; % Arrow into Data Insertion
\draw [arrow] (Insertion) -- (conv_enc) node[midway, above] {$\mathbf{v}$};
\draw [arrow] (conv_enc) -- (p_enc) node[midway, above] {$\mathbf{u}$};
\draw [arrow] (p_enc.east) -| (ch.north) node[midway, above] {$\mathbf{x}$};
\draw [arrow] (ch) |- (p_dec) node[midway, below] {$\mathbf{y}$};
\draw [arrow] (p_dec) -- (conv_dec) node[midway, above] {Metric};
\draw [arrow] (conv_dec) -- (Extraction) node[midway, above] {$\hat{\mathbf{v}}$};
\draw [arrow] (conv_dec.south) -- ++(0, -.5cm)  -| (p_dec.south)  node[pos=0.75, above, yshift=-0.33cm, xshift=-1.2cm] {Metric Request}; % Adjust position of $\hat{u}_i$
\draw [arrow] (Extraction.west) -- ++(-1cm,0) node[midway, above] {$\hat{\mathbf{d}}$}; % Arrow out of Data Extraction

\end{tikzpicture}
}
\caption{Flowchart of PAC coding scheme.}
\label{fig: flowchart}
\end{figure}

An irregular tree code (where the tree code only branches out for the data bits), polarized channels \cite{arikan2009channel}, and a tree search algorithm are the three functional blocks of a PAC coding scheme with sequential decoding, as illustrated in Fig. \ref{fig: flowchart}. The sequential decoder searches the code tree for the correct path, which corresponds to the data.
Sequential decoding was originally proposed for decoding convolutional codes (CCs) \cite{wozencraft1957sequential}, and the Fano metric is an optimal metric function used to guide the decoder to the correct path \cite{fano1963heuristic, massey1972variable}. 
For PAC codes, an optimal metric function should also consider the polarized channels \cite{moradi2021sequential, moradi2024fast}. 
Sequential decoding of polar codes using stack decoding has also been studied in \cite{niu2012stack, miloslavskaya2014sequential, trifonov2018score}.
Other works have studied successive cancellation list (SCL) decoding for PAC codes \cite{yao2021list,rowshan2021polarization}, and their weight distribution \cite{yao2021list,seyedmasoumian2022approximate, rowshan2023minimum}, among others. 
Ar{\i}kan used the design rule of Reed-Muller (RM) codes to enhance the frame error rate (FER) performance of PAC codes \cite{arikan2019sequential}. 
The RM design rule has also been studied by removing some information bits in \cite{dumer2006soft, li2014rm, moradi2023application, miloslavskaya2023design}.
The rate profile and convolutional code design of PAC codes can be viewed as an extension of polar code design, featuring dynamic frozen bits \cite{trifonov2013polar}.
Furthermore, PAC codes have also been discussed as one of the candidates for channel coding in the next-generation 6G communication \cite{wang2023road}.

Sequential decoding has a variable computation complexity, and bounds are obtained on the distribution of computations for sequential decoding of CCs on noisy discrete memoryless channels (DMCs) \cite{savage1965computation}. 
Furthermore, it has been proven that when the coding rate is below the channel cutoff rate, the computation complexity per decoded bit is bounded with high probability \cite[p.~475]{jacobs}, \cite[p.~279]{gallager1968information}. 
Conversely, for coding rates above the cutoff rate, the computation complexity for the sequential decoding of CCs grows exponentially with the block length \cite{arikan1996inequality}.
In \cite{moradi2021sequential}, the polarization effect is adapted to the sequential decoding and the Fano metric function, and simulation results demonstrated a small computation complexity per bit for the high SNR values.
Recently, in \cite{moradi2024fast}, the concept of metric function polarization is introduced, which can be leveraged for fast SCL decoding or fast sequential decoding of PAC codes. 

For a PAC\,$(N, K)$ code, let $K^-$ represent the number of data (information) bits in the first half of the code block of length $N$, and $K^+$ represent the number of data bits in the second half, where $K = K^{-} + K^{+}$ is the total number of data bits. The corresponding polarized coding rates are $R^- = K^-/(N/2)$ and $R^+ = K^+/(N/2)$, respectively, with $R = (R^{-} + R^{+})/2$ as the overall code rate. 
In \cite{moradi2023application}, it was shown that if $R^-$ exceeds the cutoff rate of the weaker polarized channel, $R_0^-$, the computation complexity for decoding the first half of the code is exponential. Similarly, if $R^+$ exceeds the cutoff rate of the stronger polarized channel, $R_0^+$, the complexity for decoding the second half of the code is exponential.
Building on this, the condition that the code rates follow the polarized cutoff rates was established as a necessary condition for bounded-complexity sequential decoding of PAC codes \cite{moradi2023application}.

In this paper, we address the converse problem of \cite{moradi2023application} by leveraging the metric polarization introduced in \cite{moradi2024fast}. 
More specifically, we characterize sufficient conditions on the PAC code rate profile to ensure a finite upper bound on the distribution of computations per data bit for sequential decoding of PAC codes. We prove that when the bias term of the metric for the first half of the code and $R^-$ are less than $R_0^-$, the distribution of computations for decoding a data bit in the first half is upper bounded by a Pareto distribution. Also, we prove the same for the second half, that if the bias term of the metric and $R^+$ are less than $R_0^+$, the distribution of computations for decoding a data bit in the second half also follows a Pareto distribution.
By continuing this analysis over the recursive polarization transform, we establish the sufficient conditions for the coding rate profile to guarantee bounded complexity based on the cutoff rate polarization. For numerical analysis, we construct a high-rate PAC\,$(1024,899)$ code based on the proposed rate profiling criteria, which demonstrate a coding gain of more than $0.75$~dB at a FER of $10^{-5}$ compared to the state-of-the-art 5G polar and LDPC codes. This is achieved while maintaining a low decoding complexity that, at high SNRs, is only within a factor of two compared to the original successive cancellation (SC) decoding of polar codes.

The rest of this paper is structured as follows.
In Section \ref{sec: preliminaries}, PAC coding scheme and Fano algorithm are briefly reviewed.
In Section \ref{sec: UpperBound} the main results are presented. 
Finally, Section \ref{sec: conclusion} concludes the paper.

%######################################

%######################################
\section{Preliminaries} \label{sec: preliminaries}

\subsection{Notation Convention}

We use bold uppercase letters for matrices and bold lowercase letters for vectors. 
For a vector $\mathbf{x} = (x_1, x_2, \ldots, x_N) \in \mathbb{F}_2^N$, $\mathbf{x}^j$ denotes the subvector $(x_1, x_2, \ldots, x_j)$, and $\mathbf{x}_i^j$ represents the subvector $(x_i, \ldots, x_j)$ for $i \leq j$.
For every subset of indices $\mathcal{A} \subset \{1, 2, \ldots, N\}$, $\mathcal{A}^c$ denotes the complement of $\mathcal{A}$, and $\mathbf{x}_{\mathcal{A}}$ represents the subvector $(x_i : i \in \mathcal{A})$. 
The notation $|\mathcal{A}|$ denotes the number of elements in the set $\mathcal{A}$.
Random variables are denoted by uppercase letters, while their realizations are denoted by lowercase letters.

%######################################
\subsection{PAC Coding}
The PAC coding scheme, as shown in Fig. \ref{fig: flowchart}, consists of three main blocks (specified by grey blocks), and is specified by $(N,K,\mathcal{A},\mathbf{p}(x))$ parameters, where $N$ and $K$ are the codeword and source word lengths, respectively, $\mathcal{A}$ is the data index set, and $\mathbf{p}(x)$ is the connection polynomial \cite{arikan2019sequential}.
A source word $\mathbf{d}= (d_1,\ldots,d_K)$ of length $K$ is generated at random and inserted into a data carrier $\mathbf{v} = (v_1, \cdots, v_{N})$, where $N = 2^n$.
After obtaining data carrier vector $\mathbf{v}$, it is encoded as $\mathbf{u} = \mathbf{v}\mathbf{T}$, where the matrix $\mathbf{T}$ is an square upper-triangular Toeplitz matrix and is constructed with polynomial $\mathbf{p}(x)$.
Then the codeword $\mathbf{u}$ is sent through a polarized channel.
A sequential decoder such as Fano algorithm is adapted to obtain an estimate vector $\hat{\mathbf{v}}$ of data carrier $\mathbf{v}$.
Finally, from $\hat{\mathbf{v}}$, an estimate $\hat{\mathbf{d}}$ of vector $\mathbf{d}$ is extracted by the data index set $\mathcal{A}$.
Obtaining the set $\mathcal{A}$ is known as the rate profiling problem.

\subsection{Fano Algorithm}
The most advantageous feature of the Fano algorithm is its ability to explore only one path at a time, which minimizes the need to store all paths and their metrics.
Due to its low memory requirements, the Fano algorithm is particularly well-suited for hardware devices with limited memory \cite{mozammel2021hardware}. 
The algorithm continues to search along a particular path as long as its metric remains high. When the metric starts to decrease substantially, the algorithm backtracks and explores alternative paths originating from earlier nodes on the previously traversed path.
The Fano algorithm begins at the root of the code tree and progresses to the child node with the highest branch metric. It proceeds to a node if its partial path metric exceeds a running threshold $T$, which is an integer multiple of a constant threshold spacing parameter $\Delta$. During a forward move, the threshold $T$ is increased in steps of size $\Delta$, up to the upper limit of the partial path metric.
If the path metric values of both child nodes are less than $T$, the Fano decoder examines the path metric of the preceding node. If the preceding node's path metric is greater than $T$, the decoder moves backward; otherwise, it reduces $T$ by $\Delta$ and attempts to proceed forward again.

For the first $j$ branches, the partial path metric is given by
\begin{equation} \label{eq: partialmetric}
    \Gamma(\mathbf{u}^j;\mathbf{y}) = \log_2 \left( \frac{P(\mathbf{y} | \mathbf{u}^{j})}{P(\mathbf{y})}\right) - \sum_{i=1}^{j}b_{i},
\end{equation}
where $\mathbf{y}$ represents the channel output, $\mathbf{u}^j$ denotes the path vector from the root of the tree to node $j$, and $b_i$ is the bias parameter \cite{moradi2021sequential}.

When evaluating a new branch in the tree, computing the branch metric is generally more efficient than calculating the partial path metric as shown in \eqref{eq: partialmetric}. To decode $u_{i}$, the decoder utilizes the channel output $\mathbf{y}$ and the preceding bits $u_{1}$ to $u_{i-1}$.
The $i$th branch metric, denoted as $\gamma(u_{i}; \mathbf{y}, \mathbf{u}^{i-1})$, allows \eqref{eq: partialmetric} to be expressed as
\begin{equation}
\Gamma(\mathbf{u}^{j}; \mathbf{y}) = \sum_{i=1}^{j} \gamma(u_{i}; \mathbf{y}, \mathbf{u}^{i-1}),
\end{equation}
where
\begin{equation}\label{eq: correct}
\gamma(u_{i}; \mathbf{y}, \mathbf{u}^{i-1})
= \log_2 \left(\frac{P(\mathbf{y}, \mathbf{u}^{i-1} \mid u_{i})}{P(\mathbf{y}, \mathbf{u}^{i-1})}\right) - b_{i}.
\end{equation}

%%%%%%%%%%%%%%%%%%%%%%%%%%%%%%%%%%%%%%%%%%%%%%%%%%%%%%%%%%%%%%%%%%%%%%%%%%%%%%%%%%%%%%%%%%%%%%%%%%%%%%%%%%%%%%%%%%%%%%%%%%%%%%%%%%%%%%

%%%%%%%%%%%%%%%%%%%%%%%%%%%%%%%%%%%%%%%%%%%%%%%%%%%%%%%%%%%%%%%%%%%%%%%%%%%%%%%%%%%%%%%%%%%%%%%%%%%%

%##########################################################################################################
\section{Upper Bound on Distribution of Computation}\label{sec: UpperBound}
In this section, we study the distribution of computation for sequential decoding of PAC codes. 
We use the standard ensemble of random linear codes for convolutional codes, as discussed in \cite[p. 206]{gallager1968information}. 
This ensemble can be represented by the pair $(\mathbf{T}, \mathbf{c})$ in the form $\mathbf{u} = \mathbf{v}\mathbf{T} + \mathbf{c}$, where $\mathbf{c}$ is a fixed, arbitrary binary vector of length $N$.

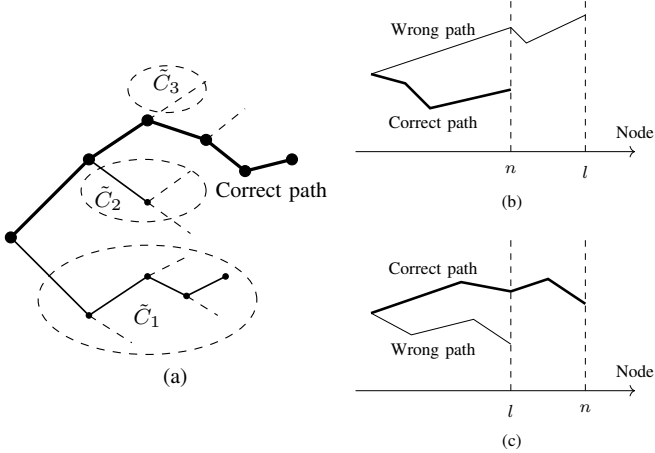
\begin{figure}[htbp]
    \centering
    \begin{tabular}{cc}
    \adjustbox{valign=c, width=0.25\textwidth}{\subfloat[\label{fig: Incorrect subsets a}]{%
          \begin{tikzpicture}

            % Draw the nodes
            \filldraw[black] (-2,0) circle (2pt);
            \filldraw[black] (-1,1) circle (2pt);
            \filldraw[black] (-.25,1.5) circle (2pt);
            \filldraw[black] (.5,1.25) circle (2pt);
            \filldraw[black] (1,.85) circle (2pt);
            \filldraw[black] (1.6,1) circle (2pt);

            % Draw the correct path
            \draw[very thick] (-2,0) -- (-1,1) -- (-.25,1.5) -- (.5,1.25) -- (1,.85) -- (1.6,1);

            % Dashed lines for incorrect path sections
            \draw[semithick] (-2,0) -- (-1,-1);
            \filldraw[black] (-1,-1) circle (1pt);
            \draw[semithick] (-1,-1) -- (-.25,-.5);
            \draw[dashed] (-1,-1) -- (-.45,-1.35);
            \draw[dashed] (-.25,-.5) -- (.25,-.25);
            \filldraw[black] (-.25,-.5) circle (1pt);
            \draw[semithick] (-.25,-.5) -- (.25,-.75);
            \filldraw[black] (.25,-.75) circle (1pt);
            \draw[semithick] (.25,-.75) -- (.75,-.5);
            \filldraw[black] (.75,-.5) circle (1pt);
            \draw[dashed] (.25,-.75) -- (.65,-1);

            \draw[dashed] (-.25,1.5) -- (.5,2);

            \draw[semithick] (-1,1) -- (-.25,.45);
            \filldraw[black] (-.25,.45) circle (1pt);
            \draw[dashed] (-.25,.45) -- (.3,.85);
            \draw[dashed] (-.25,.45) -- (.3,.05);

            \draw[dashed] (.5,1.25) -- (1,1.65);

            % Ellipses to represent incorrect subsets
            \draw[dashed] (-.25,-.8) ellipse (1.4 and 0.7);  % C1
            \draw[dashed] (-.3,0.6) ellipse (.8 and 0.4);  % C2
            \draw[dashed] (0,1.9) ellipse (.5 and 0.3);  % C3

            % Labels for the subsets
            \node at (-.25,-1) {\footnotesize $\tilde{C}_1$};
            \node at (-0.75,0.5) {\footnotesize $\tilde{C}_2$};
            \node at (0,2) {\footnotesize $\tilde{C}_3$};

            % Label for the correct path
            \node[above right] at (.5,.33) {\footnotesize Correct path};

          \end{tikzpicture}
    }} 
    \adjustbox{valign=c, width=0.23\textwidth}{\begin{tabular}{@{}c@{}}
    \subfloat[\label{fig: Incorrect subsets b}]{%
          \begin{tikzpicture}

            % Axis
            \draw[->] (-.5,0) -- (4,0) node[right, above, shift={(0,0.1)}] {\footnotesize Node};
            \draw[dashed] (2,0) -- (2,2.5);
            \draw[dashed] (3.2,0) -- (3.2,2.5);

            % Labels
            \node[below] at (2,-0.1) {\footnotesize $n$};
            \node[below] at (3.2,-0.1) {\footnotesize $l$};
            \node[above] at (3.9,0.1) {};

            % Wrong Path
            \draw[-] (-0.25,1.25) -- (2,2) -- (2.25,1.75) -- (3.2,2.2);

            % Bold path (correct path)
            \draw[very thick] (-0.25,1.25) -- (.3,1.1) -- (.7,0.7)-- (2,1.0);

            % Text labels
            \node[above] at (.75,1.7) {\footnotesize Wrong path};
            \node[below] at (.75,.7) {\footnotesize Correct path};

          \end{tikzpicture}
    } \\
    \subfloat[\label{fig: Incorrect subsets c}]{%
          \begin{tikzpicture}
            
            % Axis
            \draw[->] (-.5,0) -- (4,0) node[right, above, shift={(0,0.1)}] {\footnotesize Node};
            \draw[dashed] (2,0) -- (2,2.5);
            \draw[dashed] (3.2,0) -- (3.2,2.5);
            
            % Labels
            \node[below] at (2,-0.1) {\footnotesize $l$};
            \node[below] at (3.2,-0.1) {\footnotesize $n$};
            \node[above] at (3.9,0.1) {};
            
            % correct Path
            \draw[very thick] (-0.25,1.25) -- (1.2,1.75) -- (2,1.6) -- (2.6,1.8) -- (3.2,1.4);
            
            % (wrong path)
            \draw[-] (-0.25,1.25) -- (.4,.9) -- (1.4,1.15)-- (2,.75);
            
            % Text labels
            \node[above] at (.75,1.7) {\footnotesize Correct path};
            \node[below] at (.75,.89) {\footnotesize Wrong path};
            
            \end{tikzpicture}
          
          }
    \end{tabular}}
    \end{tabular}
    \caption{Correct path, wrong path, and incorrect subsets on code tree.}
    \label{fig: Incorrect subsets} 
\end{figure}

As shown in Fig. \ref{fig: Incorrect subsets a}, we define $\Tilde{C}_i$ as the set of extended nodes in the $i$th incorrect subtree and $C_i$ as the number of computations needed to decode the $i$th correct node (nodes on the correct path corresponding to the data). 
We have
\begin{equation}
    C_i = 1 + |\Tilde{C}_i|.
\end{equation}

In \cite[p.~475]{jacobs}, it is proven that for convolutional codes, at rates below the channel cutoff rate, the distribution of computation required to progress through each level in the tree is upper bounded by 
\begin{equation}\label{eq: UB_uncoded}
    P(C_i > L ) < AL^{-\rho},   
\end{equation}
where $A > 0$ and $\rho > 0$ are constants.
This indicates that the upper bound on the distribution of the computation for sequential decoding of convolutional codes follows a Pareto distribution.
In this paper, we generalize this result and, in our proofs, employ polarized channels that, unlike the original channel, have vector outputs.

\begin{figure}[htbp]
    \centering
    \begin{tikzpicture}[thick]

        % Input nodes
        \node at (0,0) (u1) {$\mathbf{s}_{1}^{N/2~~~}$};
        \node at (0,-1) (u2) {$\mathbf{s}_{N/2 +1}^{N}$};

        % XOR gates and dots for the first layer
        \node[draw, circle, inner sep=0pt, minimum size=3mm] at (1.5,0) (DotNo1) {};
        \node[draw, circle, fill=black, inner sep=0pt, minimum size=2pt] at (1.5,-1) (dot_u1) {};

        % Connections for the first layer
        \draw (u1) -- (DotNo1.east);
        \draw (DotNo1.east) -- (3,0) node[midway, above] {$\mathbf{x}_{1}^{N/2~~~~}$};
        \draw (u2) -- (dot_u1.west);
        \draw (dot_u1.east) -- (3,-1) node[midway, above] {$\mathbf{x}_{N/2 +1}^{N}$};

        \draw (DotNo1.north) -- (dot_u1);

        % Channel and output nodes
        \node[draw, rectangle] at (4,0) (W1) {$W^{N/2}$};
        \node[draw, rectangle] at (4,-1) (W2) {$W^{N/2}$};
        
        \node at (5.5,0) (y1) {$\mathbf{y}_{1}^{N/2~~~}$};
        \node at (5.5,-1) (y2) {$\mathbf{y}_{N/2 +1}^{N}$};

        % Connections for channel and outputs
        \draw (2,0) -- (W1.west);
        \draw (W1.east) -- (y1);
        \draw (2,-1) -- (W2.west);
        \draw (W2.east) -- (y2);

    \end{tikzpicture}
    \caption{Flowchart of a one-step polarization scheme.}
    \label{fig: OneStepPolarization}
\end{figure}
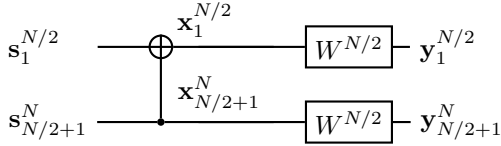

This result is relevant to PAC codes when using an uncoded system, instead of a polar code, or in other words, when PAC code is equivalent to a convolutional code. 
To demonstrate the polarization of the computation, at first we employ a one-step polarization as the inner code of the PAC code, as illustrated in Fig. \ref{fig: OneStepPolarization}.
We use the metric function for polarization, as introduced in \cite{moradi2024fast}, to analyze the polarization of the computation.

In our proof, we first examine the number of computations required by the decoder to decode the first half of the bits corresponding to the $N/2$ bad i.i.d. channels. 
If the last bit of the bad channel results in a forward move, we assume that the first half is decoded and a genie provides the correct values to decode the second half, which are inputs to the $N/2$ good i.i.d. channels.
This approach can similarly be extended to a two-step polarization process and beyond.

%######################################

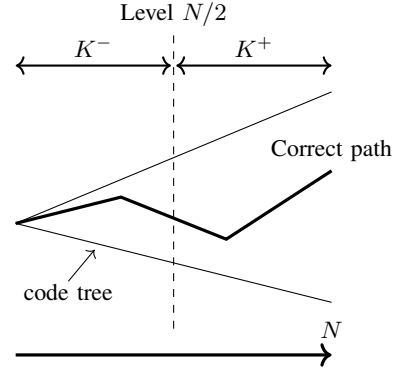
\begin{figure}[htbp] 
\centering
\small
\begin{tikzpicture}[scale=0.7]

\draw[-] (-3,3) -- (3,5.5);
\draw[-] (-3,3) -- (3,1.5);

\draw[dashed] (0,1) -- (0,6.6) node[above] {Level $N/2$};

\draw[->, very thick] (-3,.5) -- (3,.5) node[right, above, shift={(0,0.1)}] {$N$};

\draw[<->, thick] (-3,6) -- (-.05,6) node[midway, above] {$K^-$};
\draw[<->, thick] (.05,6) -- (3,6) node[midway, above] {$K^+$};

\draw[very thick] (-3,3) -- (-1,3.5) -- (1,2.7) -- (3,4)  node[above] {Correct path};

\node at (-2,1.7) {code tree};
\draw[->] (-2, 1.9) -- (-1.5, 2.5);

\end{tikzpicture}
\caption{Code tree of PAC codes after one-step polarization.} 
\label{fig: CodeTree1Step}
\end{figure}

We define the notations $\gamma^{-}(s_i) \triangleq \phi^-(s_i; y_i, y_{N/2+i}) - b^{-}$ and $\gamma^{-}(\Tilde{s}_i) \triangleq \phi^-(\Tilde{s}_i; y_i, y_{N/2+i}) - b^{-}$ to represent the $i$th branch metrics for the correct and incorrect branches, respectively, corresponding to the $i$th bad channel. 
Here, $b^{-}$ is a bias term. 
The functions $\phi^-(s_i; y_i, y_{N/2+i})$ and $\phi^-(\Tilde{s}_i; y_i, y_{N/2+i})$ \cite{moradi2024fast} are defined as
\begin{equation}
    \phi^-(s_i; y_i, y_{N/2+i}) \triangleq \log_2 \frac{P(y_i, y_{N/2+i} \mid s_i)}{P(y_i, y_{N/2+i})},
\end{equation}
and
\begin{equation}
    \phi^-(\Tilde{s}_i; y_i, y_{N/2+i}) \triangleq
    \log_2 \frac{P(y_i, y_{N/2+i} \mid \Tilde{s}_i)}{P(y_i, y_{N/2+i})}.
\end{equation}
For the realizations $\gamma^{-}(s_i)$ and $\gamma^{-}(\Tilde{s}_i)$, we use $\gamma^{-}(S_i)$ and $\gamma^{-}(\Tilde{S}_i)$ to denote the corresponding random variables. Note that $\gamma^{-}(S_i) \!\perp\!\!\!\perp \gamma^{-}(S_j)$ and $\gamma^{-}(\Tilde{S}_i) \!\perp\!\!\!\perp \gamma^{-}(\Tilde{S}_j)$ for any $i \neq j$, where $\!\perp\!\!\!\perp$ denotes statistical independence.
Fig. \ref{fig: CodeTree1Step} also illustrates the code tree of PAC codes following one-step polarization, corresponding to the representation in Fig. \ref{fig: OneStepPolarization}.

%%%%%%%%%%%%%%%%%%%%%%%%%%%%%%%%%%%%%%%%%%%%%%%%%%%%%%%%

%%%%%%%%%%%%%%%%%%%%%%%%%%%%%%%%%%%%%%%%%%%%%%%%%%%%%%%%%%%%%%%%%%%%%

%%%%%%%%%%%%%%%%%%%%%%%%%%%%%%%%%%%%%%%%%%%%%%%%%%%%%%%%%%%%%%%%%%%%%%%%%%%%%%%%%%%%%%%%%%%%%%%%%%%%%%%

%%%%%%%%%%%%%%%%%%%%%%%%%%%%%%%%%%%%%%%%%%%%%%%%

%%%%%%%%%%%%%%%%%%%%%%%%%%%%%%%%%%%%%%%%%%%%%%%%%%%%%%%%%%%%%%%%%%%%%%%%%%%%%%%%%%%%%%%%%%%%%%%%%%%%%%%%%%%%%%%%%%%%
We use $\Gamma^{-}_\text{min}$ to denote the minimum partial path metric of the correct path and $T_\text{min}^{-}$ as the minimum threshold $T$ when decoding the first half of the code. According to the Fano algorithm \cite[p.~466]{wozencraft1957sequential},
\begin{equation} \label{Gamma_min T_min}
    \Gamma^{-}_\text{min} < T^{-}_\text{min} + \Delta.
\end{equation}

We label nodes in an incorrect subset, as shown in Fig. \ref{fig: Incorrect subsets a}, using an ordered pair $(l,m)$ and denote the corresponding partial path metric by $\Tilde{\Gamma}^{-}_{l,m}$, where $l$ represents the depth of the node and $m$ is its vertical position in any arbitrary order. 
When referring to an arbitrary node at depth $l$ within an incorrect set, we use $\Tilde{\Gamma}^{-}_{l}$ to denote its partial path metric. 
The notation $\Tilde{\Gamma}^{-}_{l,m,\theta}$ is used to represent the metric of node $(l,m)$ in an incorrect set when it is visited for the $\theta$th time. 
$\Tilde{\Gamma}^{-}_{l,m}$ refers to the metric value when the node is visited for the first time. $\Tilde{\Gamma}^{-}_{l}$ represents the partial path metric at depth $l$ for a given incorrect path.

A node $(l,m)$ can be extended by the Fano algorithm if its metric satisfies the threshold condition ($\Tilde{\Gamma}^{-}_{l,m} \geq T$). 
For any specific threshold $T$, each node can be visited at most once. In any revisiting, the threshold is always lower than the previous visit by $\Delta$. 
In summary, for a given $\Tilde{\Gamma}^{-}_{l,m}$, the number of visits $\theta$ to node $(l,m)$ has an upper bound given by
\begin{equation}\label{number of visits}
    \theta < \left\lceil \frac{\Tilde{\Gamma}^{-}_{l,m} - T^{-}_{min}}{\Delta} \right\rceil \leq \frac{\Tilde{\Gamma}^{-}_{l,m} - T^{-}_{min}}{\Delta} + 1.
\end{equation}

From \eqref{Gamma_min T_min} and \eqref{number of visits}, we can conclude that a node $(l,m)$ can be visited for the $\theta$th time if
\begin{equation} \label{ith time visit}
    \Tilde{\Gamma}^{-}_{l,m} > \Gamma^{-}_{min} + (\theta-2)\Delta.
\end{equation}
Consider a binary random variable $C^{-}_{l,m,\theta}$ with a Bernoulli distribution, which takes the value 1 if node $(l,m)$ is visited for the $\theta$th time (i.e., if \eqref{ith time visit} is satisfied). 
The number of visits $C_1^{-}$ (to decode the first half of the code) has an upper bound given by
\begin{equation} \label{C_1 upper bound}
    C_1^{-} \leq 
    \sum_{l=1}^{\infty}\sum_{m}\sum_{\theta=1}^{\infty}
    C^{-}_{l,m,\theta},
\end{equation}
where $l$ denotes the depth number, $m$ denotes the vertical position, and $\theta$ denotes the number of revisits to node $(l,m)$. 
In the remainder of the paper, we aim to obtain an upper bound on the probability that $C_1^{-}$ exceeds a constant value $L$ before passing the first half of the tree.

For a binary discrete memoryless channel (B-DMC) $W$ and any $\rho \geq 0$,  
Gallager’s function associated with $W$, given an input probability distribution $q(x)$, is defined as~\cite{gallager1968information}
\begin{equation}
E_0(\rho, W) = -\log_2 \sum_{y \in \mathcal{Y}} 
\left[ \sum_{x \in \mathcal{X}} q(x) \, W(y|x)^{\frac{1}{1+\rho}} \right]^{1+\rho}.
\label{eq:E0_definition}
\end{equation}
For a random variable $X$, the moment generating function (MGF) is defined as $g(r) \triangleq \mathbb{E}[2^{rX}]$, and the semi-invariant MGF of $X$ is defined as $h(r) \triangleq \log_2 \left(\mathbb{E}[2^{rX}] \right)$. The following three lemmas will provide upper bounds on the semi-invariant MGF for the bad-channel metrics on both the correct and incorrect paths, as well as their differences.

%%%%%%%%%%%%%%%%%%%%%%%%%%%%%%%%%%%%%%%%%%%%%%%%%%%%%%%%

\begin{lemma} \label{lemma1}
    Let $h^{-}(r_0)$ denote the semi-invariant MGF of the $i$th bad-channel metric on the correct path. Then 
    \begin{equation}
    \begin{split}
        h^{-}(r_0) 
        & \triangleq \log_2 \left(\mathbb{E}\left[2^{r_0 \gamma^{-}(S_i)}\right] \right) \\
        & \leq -r_0 b^{-} - (1 + r_0) E_0\left(\frac{-r_0}{1 + r_0}, W^{-}\right),
    \end{split}
    \end{equation}
    where $r_0$ is in the interval $(-1, 0)$.
\end{lemma}

%%%%%%%%%%%%%%%%%%%%%%%%%%%%%%%%%%%%%%%%%%%%%%%

\textit{Proof:} 
The proof follows the approach of the proof of Theorem 1 in \cite{moradi2023tree}, with all intermediate steps provided to ensure the presentation is self-contained.
We first derive an upper bound for the MGF $g^{-}(r_0)$ when $-1 < r_0 < 0$.

\begin{align}
g^{-}(r_0)
&\triangleq \mathbb{E}\!\left[2^{r_0\gamma^{-}(S_i)}\right] \notag\\
&= \mathbb{E}\!\left[2^{\,r_0\log_2\!\left(\frac{P(Y_i,Y_{N/2+i}\mid S_i)}{P(Y_i,Y_{N/2+i})}\right)-r_0 b^{-}}\right] \notag\\
&= 2^{-r_0 b^{-}} \, \mathbb{E} \left[ \left( \frac{ P(Y_i, Y_{N/2+i} | S_i) }{ P(Y_i, Y_{N/2+i}) } \right)^{r_0} \right] \notag\\
&= 2^{-r_0 b^{-}} \sum_{s_i} q(s_i) \\
&~ \times \sum_{(y_i, y_{N/2+i})} 
P(y_i, y_{N/2+i} | s_i) 
\left( \frac{ P(y_i, y_{N/2+i} | s_i) }{ P(y_i, y_{N/2+i}) } \right)^{r_0} \notag\\
&= 2^{-r_0 b^{-}}
   \sum_{(y_i,y_{N/2+i})}
   \underbrace{P(y_i,y_{N/2+i})^{-r_0}}_{=\,a} \\
& ~~~~~~~~~~~~~ \times
   \underbrace{\left[\sum_{s_i} q(s_i)\,P(y_i,y_{N/2+i}\mid s_i)^{1+r_0}\right]}_{=\,b}
   .
\label{eq:sum-ab}
\end{align}

By defining $t_0 \triangleq -r_0$ with $-1 < r_0 < 0$ (hence $0 < t_0 < 1$) and applying H\"older's inequality (H)
\begin{align}
\sum_{(y_i,y_{N/2+i})} a\,b
&\le
\left(\sum_{(y_i,y_{N/2+i})} a^{\frac{1}{t_0}}\right)^{t_0}
\left(\sum_{(y_i,y_{N/2+i})} b^{\frac{1}{1-t_0}}\right)^{1-t_0},
\end{align}
we obtain
\begin{align}
& g^{-}(r_0)
\le
2^{-r_0 b^{-}}
\left[
\underbrace{
\sum_{(y_i,y_{N/2+i})} P(y_i,y_{N/2+i})
}_{=\,1}
\right]^{t_0}
\\
&  \times
\left[
\sum_{(y_i,y_{N/2+i})}
\left(
\sum_{s_i} q(s_i)\,P(y_i,y_{N/2+i}\mid s_i)^{1-t_0}
\right)^{\frac{1}{1-t_0}}
\right]^{1-t_0} \notag\\
&= 2^{-r_0 b^{-}}\;\\
&\times 2^{(1-t_0)\log_2\!
\left[
\sum_{(y_i,y_{N/2+i})}
\left(
\sum_{s_i} q(s_i)\,P(y_i,y_{N/2+i}\mid s_i)^{1-t_0}
\right)^{\frac{1}{1-t_0}}
\right]}.
\label{eq:holder-bound}
\end{align}

Recognizing the standard Gallager $E_0$ function for the minus channel $W^{-}$, the bound
\eqref{eq:holder-bound} is equivalently written as
\begin{align}
g^{-}(r_0)
&\le
2^{-r_0 b^{-}}\;
2^{-(1-t_0)\,E_0\!\left(\frac{t_0}{1-t_0},\,W^{-}\right)}
\;\\
& =\;
2^{-r_0 b^{-}-(1+r_0)\,E_0\!\left(\frac{-r_0}{1+r_0},\,W^{-}\right)}.
\end{align}

Taking the base-2 logarithm, we obtain

\begin{equation}
    h^{-}(r_0) \leq -r_0 b^{-} - (1 + r_0)E_0\left(\frac{-r_0}{1 + r_0}, W^{-}\right).
\end{equation}
\hfill\IEEEQEDhere

%%%%%%%%%%%%%%%%%%%%%%%%%%%%%%%%%%%%%%%%%%%%%%%%%%%%%%%%

\begin{lemma} \label{lemma2}
    Let $\Tilde{h}^{-}(r)$ be the semi-invariant MGF of the $i$th bad-channel metric on the wrong path. 
    Then $\Tilde{h}^{-}(r)$ is upper bounded as follows:
    \begin{equation}
        \Tilde{h}^{-}(r) \triangleq \log_2 \left(\mathbb{E}\left[2^{r\gamma^{-}(\Tilde{S_i})}\right] \right)  \leq -r b^{-} -rE_0\left(\frac{1-r}{r},W^{-}\right),
    \end{equation}
    where $r$ is in $(0,1)$ interval.
\end{lemma}

%%%%%%%%%%%%%%%%%%%%%%%%%%%%%%%%%%%%%%%%%%%%%%%
\textit{Proof:}
Consider the MGF $\Tilde{g}^{-}(r)$.
\begin{equation}
\begin{split}
 & \Tilde{g}^{-}(r) \triangleq
 \mathbb{E}\left[2^{r\gamma^{-}\left(\Tilde{S_i}\right)}\right] \\
 & = \mathbb{E}\left[2^{r\log_{2}\left(\frac{P(Y_i,Y_{N/2+i} | \Tilde{S}_i)}{P(Y_i,Y_{N/2+i})}
 \right)- rb^{-} } \right] \\
    & = \mathbb{E}\left[ \left(\frac{P(Y_i,Y_{N/2+i} | \Tilde{S}_i)}{P(Y_i,Y_{N/2+i})} \right)^{r}2^{-rb^{-}} \right]\\
    & = \sum_{\Tilde{s}_i} q(\Tilde{s}_i) \sum_{(y_i,y_{N/2+i})}
    \underbrace{\sum_{s_i} P(y_i,y_{N/2+i} | s_i)q(s_i)}
    _\text{=  $P(y_i,y_{N/2+i})$}\\
    &~~~~~ \times \left(\frac{P(y_i,y_{N/2+i} | \Tilde{s}_i)}
    {P(y_i,y_{N/2+i})} \right)^{r}2^{-rb^{-}}\\
    & = 2^{-rb^{-}} \sum_{(y_i,y_{N/2+i})} 
    \underbrace{P(y_i,y_{N/2+i})^{1-r}}_{=~b}\\
    &~~~~~~~~~~~~~~~~~~~~~~~~~~~~\times
    \underbrace{\sum_{\Tilde{s}_i}q(\Tilde{s}_i)P(y_i,y_{N/2+i} | \Tilde{s}_i)^{r}}_{=~a}.
\end{split}
\end{equation}
For $0<r<1$, and using the H\"{o}lder's inequality (H) as
\begin{equation}
    \sum ab \leq \left(\sum a^{\frac{1}{r}} \right)^{r}
    \left(\sum b^{\frac{1}{1-r}} \right)^{1-r},
\end{equation}
we obtain
\begin{equation}
\begin{split}
    & \Tilde{g}^{-}(r) 
    \leq 2^{-rb^{-}}\left[\underbrace{\sum_{(y_i,y_{N/2+i})} P(y_i,y_{N/2+i})}_\text{= 1} \right]^{1-r} \\
    &~~~~~~~~~ \left[\sum_{(y_i,y_{N/2+i})} \left[\sum_{\Tilde{s}_i} q(\Tilde{s}_i) P(y_i,y_{N/2+i} | \Tilde{s}_i)^{r} \right]^{\frac{1}{r}} \right]^{r}\\
    & = 2^{-rb^{-}} 
    2^{r\log_2\left(\sum_{(y_i,y_{N/2+i})} \left[\sum_{\Tilde{s}_i} q(\Tilde{s}_i) P(y_i,y_{N/2+i} | \Tilde{s}_i)^{r} \right]^{\frac{1}{r}} \right) }\\
    & = 2^{-r b^{-} -rE_0\left(\frac{1-r}{r},W^{-}\right)}.
\end{split}
\end{equation}
Finally, by taking the $\log_2$ function from both sides, the result becomes
\begin{equation}
        \Tilde{h}^{-}(r) = \log_2 \Tilde{g}^{-}(r)  \leq -r b^{-} -rE_0\left(\frac{1-r}{r},W^{-}\right).
\end{equation}
\hfill\IEEEQEDhere

%%%%%%%%%%%%%%%%%%%%%%%%%%%%%%%%%%%%%%%%%%%%%%%%%%%%%%%%
\begin{lemma} \label{lemma3}
    The semi-invariant MGF of the difference of the $i$th bad-channel metrics on the incorrect and correct paths has an upper bound given by
    \begin{equation}
        \log_2 \left(\mathbb{E}\left[2^{r\left(\gamma^{-}(\Tilde{S}_i) - \gamma^{-}(S_i)\right)}\right]\right)
         \leq -r b^{-} - r E_0\left(\frac{1-r}{r}, W^{-}\right),
    \end{equation}
    where \( r \) is in the interval \( (0, 1) \), and it is assumed that the bias \( b^{-} \leq \frac{E_0(\delta, W^{-})}{\delta} \) for \( 0 < \delta < 1 \).
\end{lemma}

%%%%%%%%%%%%%%%%%%%%%%%%%%%%%%%%%%%%%%%%%%%%%%%
\textit{Proof:}
Similar to the proof of Lemma \ref{lemma1}, this proof also follows the approach of the proof of Theorem 1 in \cite{moradi2023tree}, with all intermediate steps provided to ensure the presentation is self-contained.

\begin{equation*}
\begin{split}
    &\mathbb{E}\left[2^{r\left(\gamma^{-}(\Tilde{S}_i) - \gamma^{-}(S_i)\right)}\right] \\
    & = 
    \sum_{s_i}\sum_{(y_i,y_{N/2+i})}\sum_{\Tilde{s}_i}
    q(s_i)P(y_i,y_{N/2+i} | s_i)q(\Tilde{s}_i) \\
    &~~~~~~~~~~~~~~~~~~~ \times 2^{r\left(\gamma^{-}(\Tilde{s}_i) - \gamma^{-}(s_i)\right)} \\
    &= \sum_{s_i}\sum_{(y_i,y_{N/2+i})}\sum_{\Tilde{s}_i}
    q(s_i)P(y_i,y_{N/2+i} | s_i)q(\Tilde{s}_i) \\
    &~~~~~~~~~~~~~~~~~~~\times \left[\frac{P(y_i,y_{N/2+i} | \Tilde{s}_i)}{\cancel{P(y_i,y_{N/2+i})}} \right]^{r}
    \left[\frac{\cancel{P(y_i,y_{N/2+i})}}{P(y_i,y_{N/2+i} | s_i)} \right]^{r}\\
    &= \sum_{s_i}\sum_{(y_i,y_{N/2+i})}\sum_{\Tilde{s}_i}
    q(s_i)P(y_i,y_{N/2+i} | s_i)^{1-r}q(\Tilde{s}_i) \\
    & ~~~~~~~~~~~~~~~~~~~\times P(y_i,y_{N/2+i} | \Tilde{s}_i)^{r}\\
    &= \sum_{(y_i,y_{N/2+i})} 
    \underbrace{\sum_{s_i}q(s_i)P(y_i,y_{N/2+i} | s_i)^{1-r}}_\text{= $a$}  \\
    &~~~~~~~~~~~~~~~~~~~\times \underbrace{\sum_{\Tilde{s}_i}q(\Tilde{s}_i)P(y_i,y_{N/2+i} | \Tilde{s}_i)^{r}}_\text{= $b$} \\
\end{split}
\end{equation*}
\begin{equation}
\begin{split}
    & \overset{\text{H}}{\le}
    \left[\sum_{(y_i,y_{N/2+i})}\left[\sum_{s_i}q(s_i)P(y_i,y_{N/2+i} | s_i)^{1-r} \right]^{\frac{1}{1-r}} \right]^{1-r}\\
    &~~~~\times \left[\sum_{(y_i,y_{N/2+i})}\left[\sum_{\Tilde{s}_i}q(\Tilde{s}_i)P(y_i,y_{N/2+i} | \Tilde{s}_i)^{r} \right]^{\frac{1}{r}} \right]^{r}\\
    &= 2^{(1-r)\log_2\left[\sum_{(y_i,y_{N/2+i})}\left[\sum_{s_i}q(s_i)P(y_i,y_{N/2+i} | s_i)^{1-r} \right]^{\frac{1}{1-r}} \right]}\\
    &~~~~\times 2^{r\log_2\left[\sum_{(y_i,y_{N/2+i})}\left[\sum_{\Tilde{s}_i}q(\Tilde{s}_i)P(y_i,y_{N/2+i} | \Tilde{s}_i)^{r} \right]^{\frac{1}{r}} \right]}\\
    & = 2^{-(1-r)E_0(\frac{r}{1-r},W^{-})}
    2^{-rE_0(\frac{1-r}{r},W^{-})}.
\end{split}
\end{equation}
Assume that the bias $b^{-} < \frac{1-r}{r}E_0(\frac{r}{1-r},W^{-})$ for $0<r<1$ or equivalently $b^{-} < \frac{E_0(\delta,W^{-})}{\delta}$ for $0< \delta = \frac{r}{1-r}<1$.
As a result, we have
\begin{equation}
    \mathbb{E}\left[2^{r\left(\gamma^{-}(\Tilde{S}_i) - \gamma^{-}(S_i)\right)}\right] \leq
    2^{-rb^{-} -rE_0(\frac{1-r}{r},W^{-})}.
\end{equation}
We can conclude the lemma by taking monotonically increasing $\log_2$ function from both sides of the inequality.

\hfill\IEEEQEDhere

%%%%%%%%%%%%%%%%%%%%%%%%%%%%%%%%%%%%%%%%%%%%%%%%%%%%%%%%%%%%%%%%%%%%%%%%%%%%%%%%%%%%%%%%%%%%%%%%%%%%%%%%%%%%%%%%%%%%

With the help of Wald's identity, we prove the following lemma. 
The lemma states that the probability of $\Gamma^{-}_{\text{min}}$ being less than a constant value decreases exponentially fast. 
A small value of $\Gamma^{-}_{\text{min}}$ will result in a higher likelihood of the partial path metric on the incorrect path being greater than $\Gamma^{-}_{\text{min}}$, which implies that the decoder will advance further in the incorrect direction.

%%%%%%%%%%%%%%%%%%%%%%%%%%%%%%%%%%%%%%%%%%%%%%%%%%%%%%%%

\begin{lemma}\label{lemma4}
    The probability that the minimum partial path metric on the correct path is less than a constant absorbing barrier $\mu$ is upper bounded as
    \begin{equation}
        P\left(\Gamma^{-}_{\text{min}} \leq \mu \right) \leq 2^{-r_0 \mu},
    \end{equation}
    where $r_0$ is in the interval $(-1,0)$ and it is assumed that the bias $b^{-} \leq \frac{E_0(\delta,W^{-})}{\delta}$ for $0<\delta<1$.
\end{lemma}

%%%%%%%%%%%%%%%%%%%%%%%%%%%%%%%%%%%%%%%%%%%%%%%

\begin{figure}[htbp] 
\centering
	\includegraphics[width = 0.9\columnwidth]{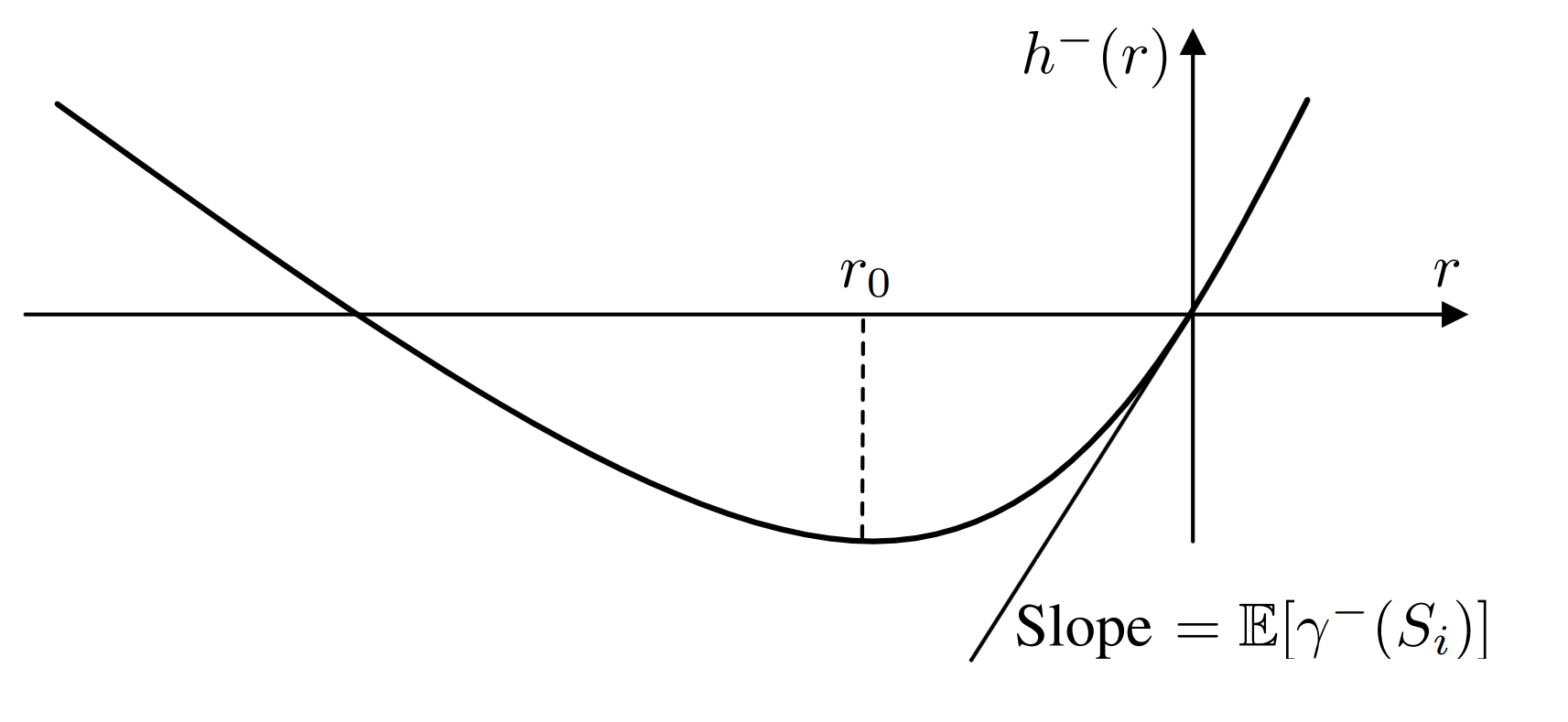}
	\caption{
    A typical plot of a semi-invariant MGF $h^{-}(r)$. The function $h^{-}(r)$ tends to $\infty$ from both sides, and its slope at $r = 0$ is $\mathbb{E}\!\left[\gamma^{-}\!\left(S_i\right)\right] = g^{-'}(0)$, where $g^{-}(r) = \mathbb{E}\!\left[2^{r\,\gamma^{-}(S_i)}\right]$ is the corresponding MGF.} 
	\label{fig: MGF}
\end{figure}

\textit{Proof:}
By taking the derivative of the semi-invariant MGF $h^{-}(r_0)$ we have
\begin{equation}
   h^{-'}(r_0) = \frac{g^{-'}(r_0)}{g^{-}(r_0)} = \frac{\mathbb{E}[\gamma^{-}(S_i)]}{g^{-}(r_0)}.
\end{equation}
As $g^{-}(0) = 1$, and at the origin we have that $h^{-'}(0) = \mathbb{E}[\gamma^{-}(S_i))]$. By taking the derivative of 
\begin{equation}
\begin{split}
    & g^{-}(r_0) = \mathbb{E}\left[2^{r_0\gamma^{-}(S_i)}\right]  
     \\
    & = \sum_{s_i}q(s_i)\sum_{(y_i,y_{N/2+i})} P(y_i,y_{N/2+i} | s_i)\\
    &~~~~~~~~~~~~~~~~~~~~~~~~~\times \left(\frac{P(y_i,y_{N/2+i} | s_i)}{P(y_i,y_{N/2+i})} \right)^{r_0} 2^{-r_0b^{-}},
\end{split}
\end{equation}
we have
\begin{equation}
\begin{split}
    & g^{-'}(r_0) = \sum_{s_i}q(s_i)\sum_{(y_i,y_{N/2+i})} P(y_i,y_{N/2+i} | s_i)\\
    &~~~~~~~\times \log_2\left(\frac{P(y_i,y_{N/2+i} | s_i)}{P(y_i,y_{N/2+i})}  \right)
    \left(\frac{P(y_i,y_{N/2+i} | s_i)}{P(y_i,y_{N/2+i})} \right)^{r_0}\\
    &~~~~~~~\times 2^{-r_0b^{-}} \ln(2)\\
    &~~~~ - \sum_{s_i}q(s_i)\sum_{(y_i,y_{N/2+i})} P(y_i,y_{N/2+i} | s_i)\\
    &~~~~~~~~~~~~~~~~~~~\times \left(\frac{P(y_i,y_{N/2+i} | s_i)}{P(y_i,y_{N/2+i})} \right)^{r_0}2^{-r_0b^{-}}b^{-} \ln(2).\\
\end{split}
\end{equation}
As a result, for $r_0 = 0$ we have
\begin{equation}
\begin{split}
    & g^{-'}(0) = \sum_{s_i}q(s_i)\sum_{(y_i,y_{N/2+i})} P(y_i,y_{N/2+i} | s_i)\\
    &~~~~~~~~~~ \log_2\left(\frac{P(y_i,y_{N/2+i} | s_i)}{P(y_i,y_{N/2+i})}  \right) \ln(2)\\
    &~~~~~~~~~~ - \sum_{s_i}q(s_i)\sum_{(y_i,y_{N/2+i})} P(y_i,y_{N/2+i} | s_i)  b^{-} \ln(2)\\
    &= I(W^{-})\ln(2) - b^{-} \ln(2).
\end{split}
\end{equation}
A random variable has a negative drift when its expectation is negative and has a positive drift when its expectation is positive. 
We see that $h^{-}(0) = \log_2(g^{-}(0)) = 0$, and
\begin{equation}
    h^{-'}(0) = \mathbb{E}\left[\gamma^{-}\left(S_i\right)\right] = g^{-'}(0) > 0 \ \ \ \ \ \text{iff} \ \ \ \ b^{-} < I(W^{-}).
\end{equation}
Using the upper bound derived in Lemma \ref{lemma1}, it can be affirmed that for an $-1<r_0<0$, the semi-invariant MGF $h^{-}(r_0)$ becomes negative if
\begin{equation}
    h^{-}(r_0) \leq -r_0 b^{-} -(1+r_0)E_0\left(\frac{-r_0}{1+r_0},W^{-}\right) < 0,
\end{equation}
where this occurs if
\begin{equation}
    b^{-} < \frac{1+r_0}{-r_0}E_0\left(\frac{-r_0}{1+r_0},W^{-}\right).
\end{equation}
So, we can conclude that $h^{-}(r_0) < 0$ if 
\begin{equation}
    b^{-} < \frac{E_0\left(\delta,W^{-}\right)}{\delta}~~~~ s.t.~~~
    \delta := \frac{-r_0}{1+r_0}, ~~~ 0< \delta < 1.
\end{equation}
Fig. \ref{fig: MGF} shows a typical behaviour of the semi-invariant MGF $h^{-}(r)$. 
Because $P\left(\gamma^{-}(S_i) > 0\right) > 0$ and $P\left(\gamma^{-}(S_i) < 0\right) > 0$, obviously we can see that $h^{-}(r) \longrightarrow \infty$  from both sides.
We are now equipped to use Wald's identity \cite[p.~434]{gallager2013stochastic} to conclude the proof.\\

% \begin{tikzpicture}[y=.2pt, x=10cm]

% \draw [thick] [->] (0,0)--(1.2,0) node[right, below] {$x$};
% \foreach \x in {0.2,0.4,0.6,0.8,1}
% \draw[thick] (\x,-1pt) -- ++(0pt,2pt) node[below] {$\x$};

% \draw [thick] [->] (0,0)--(0,10) node[above, left] {$y$};
% \foreach \y in {20,40,60}
% \draw[thick] (-1pt,\y) -- ++(2pt,0pt) node[left] {$\y$};

% \draw [domain=0:1, variable=\x, line width =1.5pt, samples=200]
% plot (\x, {-(2-(\x)^2)^8});

% \draw [domain=0:1, variable=\x, line width =1.5pt, samples=200]
% plot (\x, {-(2-(\x)^2)^8});
% \end{tikzpicture}

% \begin{figure}[htbp] 
% \centering
% \begin{tikzpicture}%[y=.2pt, x=6cm]

% \draw [thick] [->] (-5,0)--(1,0) ;

% \draw [thick] [->] (0, -1.2)--(0, 1.5) ;

% \draw [thick] [-] (-1, -2)--(.5, 1) ;
% \node at (.6, -1.7) {Slope $ = \mathbb{E}[\gamma^{-}(S_i)]$};

% \node at (-4, .8) {$h^{-}(r_0)$};

% \draw[dashed] (-1.2,-1) -- (-1.2 ,0) node [above] { $r_0$};
% % \foreach \x in {0.2,0.4,0.6,0.8,1}
% % \draw[thick] (\x,-1pt) -- ++(0pt,2pt) node[below] {$\x$};

% % \draw [thick] [->] (0,0)--(0,10) node[above, left] {$y$};
% % \foreach \y in {20,40,60}
% % \draw[thick] (-1pt,\y) -- ++(2pt,0pt) node[left] {$\y$};

% % \draw [domain=0:1, variable=\x, line width =1.5pt, samples=200]
% % plot (\x, {-(2-(\x)^2)^8});

% % \draw [domain=0:1, variable=\x, line width =1.5pt, samples=200]
% % plot (\x, {-(2-(\x)^2)^8});
% \end{tikzpicture}
% \caption{Semi-invariant MGF.} 
% \label{fig: MGF}
% \end{figure}

\textbf{Wald's identity.} 
Let $\{ \gamma^{-}(S_i); i \geq 1\}$ be i.i.d. r.v's and $h^{-}(r) = \log_2\left(\mathbb{E}[2^{r\gamma^{-}(S_i)}]\right)$ be the MGF of each $\gamma(S_i)$. Let $\Gamma^{-}_j = \sum_{i = 1}^{j} \gamma^{-}(s_i; y_i,y_{N/2+i})$ and $\Gamma^{-}_\text{min} = \inf \Gamma^{-}_j$. Then for any $r_0 < 0$ s.t. $h^{-}(r_0) \leq 0$, and any absorbing barrier $\mu$,
\begin{equation}
    P (\Gamma^{-}_{min} < \mu) \leq 2^{- r_0 \mu},
\end{equation}
where $\Gamma^{-}_{\min} = \inf \Gamma^{-}_L$ is the infimum of the partial path metric values on the correct path of the first half of the code tree.

\hfill\IEEEQEDhere 

%%%%%%%%%%%%%%%%%%%%%%%%%%%%%%%%%%%%%%%%%%%%%%%%%%%%%%%%%%%%%%%%%%%%%%%%%%%%%%%%%%%%%%%%%%%%%%%%%%%%%%%

Suppose that the correct path segment is shorter than the incorrect path segment (i.e., $n < l$) as shown in Fig. \ref{fig: Incorrect subsets b}. The following lemma is useful for providing an upper bound on the probability that the incorrect path will be extended further when $n < l$.

%##########################################################################################################

%%%%%%%%%%%%%%%%%%%%%%%%%%%%%%%%%%%%%%%%%%%%%%%%%%%%%%%
\begin{lemma} \label{lemma5}
Assuming $n < l$, we have
\begin{equation}
\begin{split}
    &P\left(\Tilde{\Gamma}^{-}_l \geq \Gamma^{-}_n + \alpha\right) 
     \leq 2^{-r \alpha} 2^{-r l\left[ E_0\left(\frac{1-r}{r},W^{-}\right) + b^{-} \right]},
\end{split}
\end{equation}
where $0<r<1$ and $b^{-} \leq \frac{E_0(\delta,W^{-})}{\delta} $ for $0<\delta<1$.
As a special case, assume that $\alpha = 0$, $r = \frac{1}{2}$, and $b^{-} = E_0(1,W^{-})$.
Then $P(\Tilde{\Gamma}^{-}_l \geq \Gamma^{-}_n)$ approaches zero exponentially with an exponent equal to $lE_0(1,W^{-})$.
\end{lemma}

%%%%%%%%%%%%%%%%%%%%%%%%%%%%%%%%%%%%%%%%%%%%%%%%

\textit{Proof:}
It is assumed that $l > n$. 
\begin{equation}
\begin{split}
    &P(\Tilde{\Gamma}^{-}_l \geq \Gamma^{-}_n + \alpha) 
     = P(2^{r \Tilde{\Gamma}^{-}_l} \geq 2^{r(\Gamma^{-}_n + \alpha)}) \\
    & \overset{\text{CB}}{\le}
     \mathbb{E} \left[ 2^{r\left[\sum_{i=1}^{l}\gamma^{-}(\Tilde{S}_i) - \sum_{i=1}^{n}\gamma^{-}(S_i) -\alpha \right]} \right] \\
    &= 2^{-r\alpha} \prod_{i=1}^{n} \mathbb{E}\left[ 2^{r(\gamma^{-}(\Tilde{S}_i) - \gamma^{-}(S_i))} \right] \prod_{i=n+1}^{l} \mathbb{E}\left[ 2^{r\gamma^{-}(\Tilde{S}_i) } \right] \\
    & \leq 2^{-r \alpha}
    2^{-r n\left[ E_0(\frac{1-r}{r},W^{-}) + b^{-} \right]} \\
    &~~~~~~~~~~~~~~~~~~~~~~~~~~~\times2^{-r (l-n)\left[ E_0(\frac{1-r}{r},W^{-}) + b^{-} \right]}\\
    & = 2^{-r \alpha} 2^{-r l\left[ E_0(\frac{1-r}{r},W^{-}) + b^{-} \right]},
\end{split}
\end{equation}
where the first inequality is by Chernoff bound (CB) and the second inequality is by Lemma \ref{lemma2} and \ref{lemma3}.

\hfill\IEEEQEDhere 

%%%%%%%%%%%%%%%%%%%%%%%%%%%%%%%%%%%%%%%%%%%%%%%%%%%%%%%%%%%%%%%%%%%%%%%%%%%%%%%%%%%%%%%%%%%%%%%%%%%%%%%

Next, assume that the correct path segment advances more than the incorrect path segment (i.e., $n \geq l$) as shown in Fig. \ref{fig: Incorrect subsets c}. The following lemma will provide an upper bound on the probability that the incorrect path will be extended further when $n \geq l$.

%%%%%%%%%%%%%%%%%%%%%%%%%%%%%%%%%%%%%%%%%%%%%%%%%%%%%%%

\begin{lemma} \label{lemma6}
Assuming that $n \geq l$, we have
\begin{equation}
    P(\Tilde{\Gamma}^{-}_l \geq \min_{n \geq l}\{\Gamma^{-}_n\} + \alpha)
    \leq 2^{r_0 \alpha} 2^{r_0 l \left( E_0\left(\frac{1+r_0}{-r_0}, W^{-}\right) + b^{-} \right)},
\end{equation}
where $r_0 \in (-1,0)$ and $b^{-} \leq \frac{E_0(\delta, W^{-})}{\delta}$ for $0 < \delta < 1$.

Similar to Lemma \ref{lemma5}, as a special case, assume that $\alpha = 0$, $r_0 = \frac{-1}{2}$, and $b^{-} = E_0(1, W^{-})$. Then, $P(\Tilde{\Gamma}^{-}_l \geq \min\{\Gamma^{-}_n\})$ goes exponentially to zero with an exponent equal to $l E_0(1, W^{-})$.
\end{lemma}

%%%%%%%%%%%%%%%%%%%%%%%%%%%%%%%%%%%%%%%%%%%%%%%%

\textit{Proof:}
\noindent Assume that $l \leq n$. We define
\begin{equation}
    \Gamma_{l+1}^{n} \triangleq \sum_{i=l+1}^{n}\gamma^{-}(S_i),
\end{equation}
for $n > l$, and $\Gamma_{l+1}^{n} = 0$ for $n = l$. In this case ($l \leq n$), we have
\begin{equation}
    \Gamma^{-}_\text{min} = \Gamma^{-}_l + \inf_{\forall n \geq l} \{\Gamma_{l+1}^n\}.
\end{equation}
Thus,
\begin{equation}
\begin{split}
    &P(\Tilde{\Gamma}^{-}_l \geq \Gamma^{-}_\text{min} + \alpha) \\
    &= P\left[\Tilde{\Gamma}^{-}_l \geq \Gamma^{-}_l + \inf_{\forall n \geq l} \{\Gamma_{l+1}^n\} + \alpha \right] \\
    &= P\left[\Tilde{\Gamma}^{-}_l - \Gamma^{-}_l - \alpha - \inf_{\forall n \geq l} \{\Gamma_{l+1}^n\} \geq 0 \right] \\
    &= \sum_{\mu} P(\Tilde{\Gamma}^{-}_l - \Gamma^{-}_l - \alpha = \mu) P(\inf_{\forall n \geq l} \{\Gamma_{l+1}^n\} \leq \mu) \\
    &\leq \sum_{\mu} P(\Tilde{\Gamma}^{-}_l - \Gamma^{-}_l - \alpha = \mu) 2^{-r_0 \mu} \\
    &= \mathbb{E}\left[2^{-r_0( \Tilde{\Gamma}^{-}_l - \Gamma^{-}_l - \alpha )} \right] \\
    &= 2^{r_0 \alpha} \prod_{i=1}^{l} \mathbb{E}\left[2^{-r_0(\gamma^{-}(\Tilde{S}_i) - \gamma^{-}(S_i))} \right] \\
    &\leq 2^{r_0 \alpha} \prod_{i=1}^{l} 2^{r_0 [E_0(\frac{1+r_0}{-r_0},W^{-}) + b^{-}]} \\
    &= 2^{r_0 \alpha} 2^{r_0 l [E_0(\frac{1+r_0}{-r_0}, W^{-}) + b^{-}]},
\end{split}
\end{equation}
where the first inequality is an application of Wald's identity, and the second inequality follows from Lemma \ref{lemma3}.

\hfill\IEEEQEDhere

%%%%%%%%%%%%%%%%%%%%%%%%%%%%%%%%%%%%%%%%%%%%%%%%%%%%%%%%%%%%%%%%%%%%%%%%%%%%%%%%%%%%%%%%%%%%%%%%%%%%%%%
Whenever the partial path metric of any incorrect path at a given depth is above $\Gamma_\text{min}$, the wrong direction has the chance to continue further. The following theorem provides an upper bound on the probability that this can occur.

%%%%%%%%%%%%%%%%%%%%%%%%%%%%%%%%%%%%%%%%%%%%%%%%%%%%%%%%
\begin{theorem}\label{theorem 1}
The probability that the partial path metric of the incorrect path is greater than or equal to the minimum partial path metric of the correct path by a constant $\alpha$ is upper bounded as
\begin{equation}
\begin{split}
    P\left( \Tilde{\Gamma}^{-}_l \geq \Gamma^{-}_\text{min} + \alpha \right) 
    \leq (l+1) 2^{-r\alpha} 2^{-r l\left[ E_0\left(\frac{1-r}{r},W^{-}\right) + b^{-} \right]},
\end{split}
\end{equation}
where it is assumed that $r \in (0,1)$ and $b^{-} < \frac{E_0(\delta,W^{-})}{\delta}$ for $0 < \delta < 1$. Similarly, for $\alpha = 0$, $r = \frac{1}{2}$, and $b^{-} = E_0(1,W^{-})$, the upper bound exponent is $lE_0(1,W^{-})$, and the chance of advancing more than $l$ steps in the wrong path has an upper bound which is a linear function of $l$.
\end{theorem}

%%%%%%%%%%%%%%%%%%%%%%%%%%%%%%%%%%%%%%%%%%%%%%%

\textit{Proof:}
\begin{equation}
\begin{split}
    P\left( \Tilde{\Gamma}^{-}_l \geq \Gamma^{-}_\text{min} + \alpha \right) 
    &\leq \sum_{n = 0}^{l-1}P\left( \Tilde{\Gamma}^{-}_l \geq \Gamma^{-}_{n} + \alpha \right)\\
    &\quad + 
    P\left( \Tilde{\Gamma}^{-}_l \geq \Gamma^{-}_{l} + \inf_{\forall n \geq l} \{\Gamma_{l+1}^n\} + \alpha \right)\\
    &\leq \sum_{n = 0}^{l-1} 2^{-r \alpha} 2^{-r l \left[ E_0\left(\frac{1-r}{r},W^{-}\right) + b^{-} \right]} \\
    &\quad + 2^{r_0 \alpha} 2^{r_0 l \left[ E_0\left(\frac{1+r_0}{-r_0},W^{-}\right) + b^{-} \right]} \\
    &= (l+1) 2^{-r \alpha} 2^{-r l \left[ E_0\left(\frac{1-r}{r},W^{-}\right) + b^{-} \right]},
\end{split}
\end{equation}
where the first inequality is by the definition of $\Gamma^{-}_\text{min}$ and Boole's inequality. The second inequality is by Lemma \ref{lemma5} and Lemma \ref{lemma6}. The last equality is obtained by assuming $r = -r_0$.

\hfill\IEEEQEDhere

%%%%%%%%%%%%%%%%%%%%%%%%%%%%%%%%%%%%%%%%%%%%%%%%%%%%%%%%%%%%%%%%%%%%%%%%%%%%%%%%%%%%%%%%%%%%%%%%%%%%%%%

Suppose that the decoder is in the $(l,m)$th node of the decoding tree.
We define partial rate as $R^{-}_l = \frac{\lambda^{-}_l}{l}$, where $\lambda^{-}_{l}$ is the number of times the irregular tree code branches up to the depth $l$ on the first half of the code tree.
We can see that $2^{\lambda^{-}_l} = 2^{lR^{-}_l}$ is an upper bound on the number of incorrect nodes at depth $l$ of the tree. 

Furthermore, assume that 
\begin{equation} \label{partial rate}
    R^{-}_l \leq r\left(  E_0(\frac{1-r}{r},W^{-})+b^{-}    \right)- \epsilon,
\end{equation}
where $\epsilon$ is a small positive number. 
We would like to mention that by substituting the parameters as $r = \frac{1}{2}$ and $b^{-} = E_0(1,W_N^{-})$, the inequality reduces to
\begin{equation}
     R^{-}_l \leq   E_0(1,W^{-}) -\epsilon .
\end{equation}
%the sum of the first $l$ bit-channel cutoff rates are less than the number of the information bits in the first $l$ bits as

With the conditions mentioned above, the following theorem gives an upper bound for the $\mathbb{E}[C_1^{-}],$ which corresponds to the average number of computations needed to decode the first bit. The average is over data sequence, the channel noise, and the ensemble of the PAC codes.

By (\ref{C_1 upper bound}), we have
\begin{equation} \label{expected C0}
\begin{split}
    & \mathbb{E}[C_1^{-}] \leq \sum_{l=1}^{\infty}\sum_{m}\sum_{\theta=1}^{\infty} \mathbb{E}[C^{-}_{l,m,\theta}] \\
    & = \sum_{l=1}^{\infty}\sum_{m}\sum_{\theta=1}^{\infty}
    P(\Tilde{\Gamma}^{-}_{l,m} > \Gamma^{-}_{min} + (\theta - 2)\Delta),
\end{split}
\end{equation}
where $C^{-}_{l,m,\theta}$ has a Bernoulli distribution with probability of being one equal to $P(\Tilde{\Gamma}^{-}_{l,m} > \Gamma^{-}_{min} + (\theta - 2)\Delta)$.

%%%%%%%%%%%%%%%%%%%%%%%%%%%%%%%%%%%%%%%%%%%%%%%%%%%%%%%%
\begin{theorem}\label{theorem 2}
The value $\Delta = \frac{1}{r}$ minimizes the upper bound of $\mathbb{E}[C_1^{-}]$ to
\begin{equation}
    \mathbb{E}[C_1^{-}] \leq  
    \frac{4}{(1-2^{-\epsilon})^2},
\end{equation}
where $\epsilon$ satisfies (\ref{partial rate}).
\end{theorem}

%%%%%%%%%%%%%%%%%%%%%%%%%%%%%%%%%%%%%%%%%%%%%%%

\textit{Proof:}
\begin{equation}
\begin{split}
    & \mathbb{E}[C_1^{-}] \leq \sum_{l=1}^{\infty}\sum_{m}\sum_{\theta=1}^{\infty} \mathbb{E}[C^{-}_{l,m,\theta}] \\
    & = \sum_{l=1}^{\infty}\sum_{m}\sum_{\theta=1}^{\infty}
    P(\Tilde{\Gamma}^{-}_{l,m} > \Gamma^{-}_\text{min} + (\theta - 2)\Delta),\\
    &\leq \sum_{l=1}^{\infty}\sum_{m}\sum_{\theta=1}^{\infty}
    (l+1) 2^{-r\alpha} 2^{-rl\left[ E_0(\frac{1-r}{r},W^{-}) + b^{-} \right] } \\
    &= \sum_{l=1}^{\infty}\sum_{\theta=1}^{\infty}
    (l+1) 2^{lR^{-}_{l}} 2^{-r\alpha}2^{-r l \left[ E_0(\frac{1-r}{r},W^{-}) + b^{-} \right] } \\
    & \leq \sum_{\theta=1}^{\infty}2^{-r\alpha} \sum_{l=1}^{\infty}
    (l+1)2^{lR^{-}_{l}}  2^{-(lR^{-}_{l} + l\epsilon) } \\
    &= \sum_{\theta=1}^{\infty}2^{-r\alpha} \sum_{l=1}^{\infty}
    (l+1) \left[ 2^{-\epsilon }\right]^{l} 
    \leq \sum_{\theta=1}^{\infty}2^{-r\alpha} \frac{1}{(1-2^{-\epsilon})^2} \\
    & = \sum_{\theta=1}^{\infty}2^{-r(\theta - 2)\Delta} 
    \frac{1}{(1-2^{-\epsilon})^2}
    = \frac{2^{r\Delta}}{1-2^{-r\Delta}}\frac{1}{(1-2^{-\epsilon})^2},
\end{split}
\end{equation}
for $\alpha = (\theta - 2)\Delta$.
The second inequality is by Theorem \ref{theorem 1}. 
The third inequality is by  (\ref{partial rate}) and the upper bound on the number of incorrect nodes at depth $l$ of the decoding tree. 
The infinite sigma on the depth of the tree $l$ is convergent if and only if $\epsilon$ is positive, as we assumed it.

The value $\Delta = \frac{1}{r}$ minimizes the upper bound. 
So we have
\begin{equation}
    \mathbb{E}[C_1^{-}] \leq  
    \frac{4}{(1-2^{-\epsilon})^2}.
\end{equation}

\hfill\IEEEQEDhere 

%%%%%%%%%%%%%%%%%%%%%%%%%%%%%%%%%%%%%%%%%%%%%%%%%%%%%%%%%%%%%%%%%%%%%%%%%%%%%%%%%%%%%%%%%%%%%%%%%%%%%%%

Ultimately, the following theorem provides an upper bound on the CCDF of the bad-channel computations.

%%%%%%%%%%%%%%%%%%%%%%%%%%%%%%%%%%%%%%%%%%%%%%%%%%%%%%%%
\begin{theorem}\label{theorem 3}
Assume that $b^{-} \leq \frac{E_0(\delta, W^{-})}{\delta}$ for $0<\delta<1$, and 
\begin{equation} 
    R^{-}_{l} \leq \frac{r}{\beta} \left( E_0\left(\frac{1-r}{r}, W^{-}\right) + b^{-}\right) - \epsilon  ,
\end{equation}
where $0<r<1$ and $\epsilon > 0$.
Then the probability that the number of computations required to decode the $n$th bit of the first half of the decoding tree, $C_n^{-}$, exceeds a constant value $L$ has an upper bound given by the Pareto distribution as 
\begin{equation}
    P(C_n^{-} \geq L) \leq \frac{\mathbb{E}[C_n^{-\beta}]}{L^{\beta}}
    \leq \left(\frac{4}{L(1-2^{-\epsilon/\beta})^2}\right)^{\beta},
\end{equation}
where $\beta > 1$. 
\end{theorem}

%%%%%%%%%%%%%%%%%%%%%%%%%%%%%%%%%%%%%%%%%%%%%%%

\textit{Proof:}
For $\beta > 1$, probabilities $Q_j$, and a set of nonnegative numbers $a_{jk}$, the Minkowski inequality (MI) is given by
\begin{equation}
    \left[\sum_j Q_j \left(\sum_k a_{jk} \right)^{\beta} \right]^{1/\beta}
    \leq \sum_{k} \left(\sum_j Q_j a_{jk}^{\beta} \right)^{1/\beta}.
\end{equation}
By using the Minkowski inequality and \eqref{expected C0}, we get
\begin{equation}
\begin{split}
    (\mathbb{E}[C_1^{-\beta}])^{1/\beta}
    &\leq \left[ \mathbb{E}\left(\left[\sum_{l=1}^{\infty}\sum_{m}\sum_{\theta=1}^{\infty} C^{-}_{l,m,\theta} \right]^{\beta}\right) \right]^{1/\beta}\\
    &\overset{\text{MI}}{\le} 
    \sum_{l=1}^{\infty}\sum_{m}\sum_{\theta=1}^{\infty} \left( \mathbb{E} \left[ (C^{-}_{l,m,\theta})^{\beta} \right]\right)^{1/\beta}.
\end{split}
\end{equation}

Since $C^{-}_{l,m,\theta}$ is a random variable with a Bernoulli distribution, we have $(C^{-}_{l,m,\theta})^{\beta} = C^{-}_{l,m,\theta}$. Similar to the proof of the previous theorem and assuming $\alpha = (\theta-2)\Delta$, we have
\begin{equation}
\begin{split}
    (\mathbb{E}[C_1^{-\beta}])&^{1/\beta}
    \leq \sum_{l=1}^{\infty}\sum_{m}\sum_{\theta=1}^{\infty} \left( \mathbb{E} \left[ (C^{-}_{l,m,\theta})^{\beta} \right]\right)^{1/\beta}\\
    &= \sum_{l=1}^{\infty}\sum_{m}\sum_{\theta=1}^{\infty} P\left[ \Tilde{\Gamma}^{-}_l \geq \Gamma^{-}_\text{min} + (\theta - 2)\Delta \right]^{1/\beta} \\
    &\leq \sum_{l=1}^{\infty}\sum_{m}\sum_{\theta=1}^{\infty} (l+1)^{1/\beta} 2^{-\frac{r\alpha}{\beta} } 2^{-\frac{r}{\beta} l\left[ E_0\left(\frac{1-r}{r}, W^{-}\right) + b^{-} \right] } \\
    & = \sum_{l=1}^{\infty}\sum_{\theta=1}^{\infty} (l+1)^{1/\beta}2^{lR^{-}_{l}} 2^{-\frac{r\alpha}{\beta} } 2^{-\frac{r}{\beta} l\left[ E_0\left(\frac{1-r}{r}, W^{-}\right) + b^{-} \right] } \\    
    &\leq \sum_{\theta=1}^{\infty} 2^{-\frac{r\alpha}{\beta}} \sum_{l=1}^{\infty} (l+1)^{1/\beta} 2^{lR^{-}_{l}} 2^{-(lR^{-}_{l} + l\epsilon)} \\
    &\leq \sum_{\theta=1}^{\infty} 2^{-\frac{r\alpha}{\beta}} \sum_{l=1}^{\infty} (l+1)^{1/\beta} \left(2^{-\epsilon}\right)^{l} \\
    &\leq \sum_{\theta=1}^{\infty} 2^{-\frac{r\alpha}{\beta}} \sum_{l=1}^{\infty} (l+1) \left(2^{-\epsilon}\right)^{l} \\
    &\leq \sum_{\theta=1}^{\infty} 2^{-\frac{r\alpha}{\beta}} \frac{1}{(1 - 2^{-\epsilon})^2} \\
    &= \sum_{\theta=1}^{\infty} 2^{-\frac{r(\theta - 2)\Delta}{\beta}} \frac{1}{(1 - 2^{-\epsilon})^2} \\
    &= \frac{2^{\frac{r\Delta}{\beta}}}{1 - 2^{-\frac{r\Delta}{\beta}}} \frac{1}{(1 - 2^{-\epsilon})^2}.
\end{split}
\end{equation}

The value of $\Delta = \frac{\beta}{r}$ (threshold spacing) will minimize the upper bound, and we obtain
\begin{equation}
     (\mathbb{E}[C_1^{-\beta}])^{1/\beta} \leq \frac{4}{(1 - 2^{-\epsilon})^2}.
\end{equation}

Due to the symmetry of the problem, the same bound is valid for any node. Finally, using the generalized Chebyshev inequality, we have
\begin{equation}
    P(C_n^{-} \geq L) \leq \frac{\mathbb{E}[C_n^{-\beta}]}{L^{\beta}}.
\end{equation}

\hfill\IEEEQEDhere

%%%%%%%%%%%%%%%%%%%%%%%%%%%%%%%%%%%%%%%%%%%%%%%%%%%%%%%%%%%%%%%%%%%%%%%%%%%%%%%%%%%%%%%%%%%%%%%%%%%%%%%

By assuming that a genie provides the correct value of $s_i$, we define 
$$\gamma^{+}(s_{N/2+i}) \triangleq \phi^{+}(s_{N/2+i}; y_i, y_{N/2+i},s_i) - b^{+}$$
and 
$$\gamma^{+}(\Tilde{s}_{N/2+i}) \triangleq \phi^{+}(\Tilde{s}_{N/2+i}; y_i, y_{N/2+i},s_i) - b^{+}$$
to represent the $i$th branch metrics for the correct and incorrect branches, respectively, corresponding to the $i$th good channel. 
Here, $b^{+}$ is a bias term. 
The functions $\phi^{+}(s_{N/2+i}; y_i, y_{N/2+i},s_i)$ and $\phi^{+}(\Tilde{s}_{N/2+i}; y_i, y_{N/2+i},s_i)$ \cite{moradi2024fast} are defined as
\begin{equation}
    \phi^{+}(s_{N/2+i}; y_i, y_{N/2+i},s_i) \triangleq \log_2 \frac{P(y_i, y_{N/2+i},s_i \mid s_{N/2+i})}{P(y_i, y_{N/2+i})},
\end{equation}
and
\begin{equation}
    \phi^{+}(\Tilde{s}_{N/2+i}; y_i, y_{N/2+i},s_i) \triangleq
    \log_2 \frac{P(y_i, y_{N/2+i},s_i \mid \Tilde{s}_{N/2+i})}{P(y_i, y_{N/2+i},s_i)}.
\end{equation}
For the realizations $\gamma^{+}(s_{N/2+i})$ and $\gamma^{+}(\Tilde{s}_{N/2+i})$, we use $\gamma^{+}(S_{N/2+i})$ and $\gamma^{+}(\Tilde{S}_{N/2+i})$ to denote the corresponding random variables. 
Note that $\gamma^{+}(S_{N/2+i}) \!\perp\!\!\!\perp \gamma^{+}(S_{N/2+j})$ and $\gamma^{+}(\Tilde{S}_{N/2+i}) \!\perp\!\!\!\perp \gamma^{+}(\Tilde{S}_{N/2+j})$ for any $i \neq j$, where $\!\perp\!\!\!\perp$ denotes statistical independence.
Assume that with the assistance of a genie, the first half of the code has been decoded, and now the decoder aims to decode the second half. 
Similarly, we can prove the following theorem.

%%%%%%%%%%%%%%%%%%%%%%%%%%%%%%%%%%%%%%%%%%%%%%%%%%%%%%%%
\begin{theorem}\label{theorem 4}
Assume that \( b^{+} \leq \frac{E_0(\delta, W^{+})}{\delta} \) for \( 0 < \delta < 1 \), and 
\begin{equation}
    R^{+}_{l} \leq \frac{r}{\beta} \left( E_0\left(\frac{1 - r}{r}, W^{+}\right) + b^{+}\right) - \epsilon,
\end{equation}
where \( 0 < r < 1 \) and \( \epsilon > 0 \).
Then the probability that the number of computations required to decode the \( n \)th bit of the second half of the decoding tree, \( C_n^{+} \), exceeds a constant value \( L \) is upper bounded by a Pareto distribution as
\begin{equation}
    P(C_n^{+} \geq L) \leq \frac{\mathbb{E}[{C_n^{+}}^{\beta}]}{L^{\beta}} \leq \left(\frac{4}{L(1 - 2^{-\epsilon/\beta})^2}\right)^{\beta},
\end{equation}
where \( \beta > 1 \). 
\end{theorem}

%%%%%%%%%%%%%%%%%%%%%%%%%%%%%%%%%%%%%%%%%%%%%%%
Note that by assuming \( r = \frac{1}{2} \) and \( \beta = 1 \), the condition becomes \( R^{+}_l \leq E_0(1, W^+) - \epsilon = R^{+}_{0} - \epsilon \).
This also suggests setting \(\Delta = 2\), which is the desired value for the simulations \cite{moradi2021sequential}.
Similarly, the polarization operation can be repeated recursively. This procedure implies that the PAC code's polarized rates and biases should be lower than the polarized cutoff rate to achieve computations with a Pareto distribution.

For a general notation, consider a convolutional code with output $\mathbf{r} = \mathbf{v}\mathbf{T} + \mathbf{c}$, where the polar mapper undergoes $k$ steps of polarization. Define
\begin{equation*}
    \prescript{}{i}{\mathbf{r}} \triangleq \left(r_i, r_{N/2^k + i}, \ldots, r_{(2^k-1)N/2^k + i}\right)
\end{equation*}
and its corresponding channel output as 
\begin{equation*}
     \prescript{}{i}{\mathbf{y}} \triangleq \left(y_i, y_{N/2^k + i}, \ldots, y_{(2^k-1)N/2^k + i}\right)
\end{equation*}
for $i$ ranging from 1 to $N/2^k$.
By employing $k$ steps of polarization, we have $2^k$ type of $N/2^k$ i.i.d. channels 
$W^{\{-,+\}^{k}}(\prescript{}{i}{r}_j; \prescript{}{i}{\mathbf{y}}, \prescript{}{i}{\mathbf{r}}_1^{j-1}) $ 
from the original $N$ i.i.d. channels $W$. 
Assume the decoder has obtained $\prescript{}{i}{\mathbf{r}}_1^{j-1}$. 

We define the notation 
$$\gamma^{\{-,+\}^{k}}(\prescript{}{i}{r}_j) \triangleq
\phi^{\{-,+\}^{k}}(\prescript{}{i}{r}_j; 
     \prescript{}{i}{\mathbf{y}}, \prescript{}{i}{\mathbf{r}}_1^{j-1}) - b^{\{-,+\}^{k}}
$$
and 
$$\gamma^{\{-,+\}^{k}}(\prescript{}{i}{\Tilde{r}}_j) \triangleq
\phi^{\{-,+\}^{k}}(\prescript{}{i}{\Tilde{r}}_j; 
     \prescript{}{i}{\mathbf{y}}, \prescript{}{i}{\mathbf{r}}_1^{j-1}) - b^{\{-,+\}^{k}},
$$
where $i$ ranges from 1 to $N/2^k$, corresponding to the $i$-th channel of $W^{\{-,+\}^{k}}(\prescript{}{i}{r}_j; 
\prescript{}{i}{\mathbf{y}}, \prescript{}{i}{\mathbf{r}}_1^{j-1}) $ and $b^{\{-,+\}^{k}}$ are the bias terms. 
Also, the function $\phi^{\{-,+\}^{k}}(\prescript{}{i}{r}_j; 
\prescript{}{i}{\mathbf{y}}, \prescript{}{i}{\mathbf{r}}_1^{j-1})$ 
is defined as
\begin{equation}
\phi^{\{-,+\}^{k}}(\prescript{}{i}{r}_j; 
     \prescript{}{i}{\mathbf{y}}, \prescript{}{i}{\mathbf{r}}_1^{j-1}) 
    \triangleq \log_2 \left( \frac{P(\prescript{}{i}{\mathbf{y}}, \prescript{}{i}{\mathbf{r}}_1^{j-1} \mid \prescript{}{i}{r}_j)}{P(\prescript{}{i}{\mathbf{y}}, \prescript{}{i}{\mathbf{r}}_1^{j-1})} \right),
\end{equation}

Similarly, the metric of the wrong branch is defined as
\begin{equation}
\phi^{\{-,+\}^{k}}(\prescript{}{i}{\Tilde{r}}_j; 
     \prescript{}{i}{\mathbf{y}}, \prescript{}{i}{\mathbf{r}}_1^{j-1}) 
    \triangleq \log_2 \left( \frac{P(\prescript{}{i}{\mathbf{y}}, \prescript{}{i}{\mathbf{r}}_1^{j-1} \mid \prescript{}{i}{\Tilde{r}}_j)}{P(\prescript{}{i}{\mathbf{y}}, \prescript{}{i}{\mathbf{r}}_1^{j-1})} \right).
\end{equation}
For the obtained $N/2^{k}$ i.i.d. channels, we can prove the following theorem.
%%%%%%%%%%%%%%%%%%%%%%%%%%%%%%%%%%%%%%%%%%%%%%%%%%%%%%%%
\begin{theorem}\label{theorem 5}
Assume that, with the help of a genie, the previous chunks of the decoding tree, each with a length of \( \frac{N}{2^k} \), have been decoded, and the decoder is now tasked with decoding the \( a \)th chunk.
Suppose
\[
b^{\{-,+\}^k} \leq \frac{E_0(\delta, W^{\{-,+\}^k})}{\delta}
\]
for \( 0 < \delta < 1 \), and
\begin{equation}
    R^{\{-,+\}^k}_{l} \leq \frac{r}{\beta} \left( E_0\left(\frac{1 - r}{r}, W^{\{-,+\}^k}\right) + b^{\{-,+\}^k}\right) - \epsilon,
\end{equation}
where \( 0 < r < 1 \) and \( \epsilon > 0 \).
Then the probability that the number of computations required to decode the \( n \)th bit of this chunk in the decoding tree, \( C_n^{\{-,+\}^k} \), exceeds a constant \( L \) is upper-bounded by a Pareto distribution as
\begin{equation}
    P(C_n^{\{-,+\}^k} \geq L) \leq \frac{\mathbb{E}\left[{C_n^{\{-,+\}^k}}^\beta\right]}{L^\beta} \leq \left(\frac{4}{L(1 - 2^{-\epsilon/\beta})^2}\right)^\beta,
\end{equation}
where \( \beta > 1 \). 
\end{theorem}
Note that by assuming \( r = \frac{1}{2} \) and \( \beta = 1 \), the condition simplifies to
\begin{equation}\label{eq: cutoffConstraint delta2}
    R^{\{-,+\}^k}_l \leq E_0(1, W^{\{-,+\}^k}) - \epsilon = R^{\{-,+\}^k}_0 - \epsilon.
\end{equation}

This theorem presents the converse of the results in \cite{moradi2023application}. 
As outlined in \cite{moradi2023application}, when \( R^{\{-,+\}^k}_{\frac{N}{2^k}} > R^{\{-,+\}^k}_0 \), the computation required to decode the corresponding chunk grows exponentially with \( \frac{N}{2^k} \).
Conversely, based on Theorem \ref{theorem 5}, if \( R^{\{-,+\}^k}_{\frac{N}{2^k}} < R^{\{-,+\}^k}_0 \), the probability that the number of computations required to decode the corresponding chunk exceeds \( L \) is bounded by a constant.
Based on these results, we adopt the rate-profile design from \cite{moradi2023application} to achieve a high reliability code with a low-complexity decoder.
%%%%%%%%%%%%%%%%%%%%%%%%%%%%%%%%%%%%%%%%%%%%%%%

%%%%%%%%%%%%%%%%%%%%%%%%%%%%%%%%%%%%%%%%%%%%%%%%%%%%%%%%%%%%%%%%%%%%%%%%%%%%%%%%%%%%%%%%%%%%%%%%%%%%%%%%%%%%%%%%%%%%

\section{A Rate-Profile Design Method and Numerical Results}
As our theoretical results suggest, we set $\Delta = 2$ for all of our numerical results when there is no constraint on the amount of search in the Fano decoding.

For a polarization step of \( k = 1 \), the channel cutoff rate lies strictly below the channel capacity, and a convolutional code with Fano decoding can only achieve the original channel cutoff rate with a low computational complexity \cite{jacobs}.  
With higher polarization steps, the average of the polarized channel cutoff rates approaches the channel capacity, allowing Fano decoding to operate closer to the channel capacity while maintaining low computational complexity.  
When \( k = n \) and \( N = 2^n \) tends to infinity, the channel cutoff rate polarizes almost surely \cite{arikan2009channel}, and a rate profile can allocate data to positions whose polarized cutoff rate equals 1.  
In the numerical results presented in this paper, we designed the rate profiles for the case \( 2^{n-k} = 8 \).
The computation polarization described in Theorem \ref{theorem 5} imposes a constraint on the rate profile of the code. 
Building on this, we propose a rate profile design. 
Specifically, we detail an algorithm to construct the rate profile for a PAC\,$(1024, 899)$ code, adhering to these constraints.
We begin with an RM$(1024, 968)$ code, which has a minimum distance of 8. 
We assume a polarization step of \( k = 7 \), resulting in 128 chunks, each of length 8. For a given target \( E_b/N_0 \), we freeze the data bits of the RM$(1024, 968)$ code to ensure that the polarized rate profile constraint is satisfied.
At a target \( E_b/N_0 = 3~\text{dB} \), we must freeze at least 93 of the data bits from the RM$(1024, 968)$ code to satisfy the constraint imposed by the polarization cutoff rate, resulting in a $(1024, 875)$ code. This indicates that either the target \( E_b/N_0 \) must be increased, or a lower minimum distance code must be used to construct a PAC\,$(1024, 899)$ code.
We increase the target $E_b/N_0$ to $3.6~\text{dB}$ and after fulfilling the polarization cutoff rate constraint, we obtain 909 data bits.
We then freeze the 10 data bits, corresponding to the last rows of the generator matrix with weight 8 of the obtained (1024,909) code, to construct a PAC\,$(1024, 899)$ code.
We guess that freezing the bits corresponding to the last rows enhances the upper bound of FER performance more effectively than freezing other bits of equivalent weight, analogous to zero padding in convolutional codes, and this requires further investigation.
In our numerical example, we also include the FER of the PAC codes obtained by freezing the 10 data bits corresponding to the upper rows (PAC-Fano, FS), which results in significantly worse error-correction performance. This is analogous to the case in convolutional codes where zero padding is not used.
As a related work on obtaining a rate profile with a bounded search complexity (moderate list size) and improving the rate profile of 5G polar codes, the rate profiles in \cite{cocskun2022information} mainly freeze the bits corresponding to the lower rows, while using the bits corresponding to the upper rows as information bits.

\begin{figure}[htbp] 
\centering
	\includegraphics[width = \columnwidth]{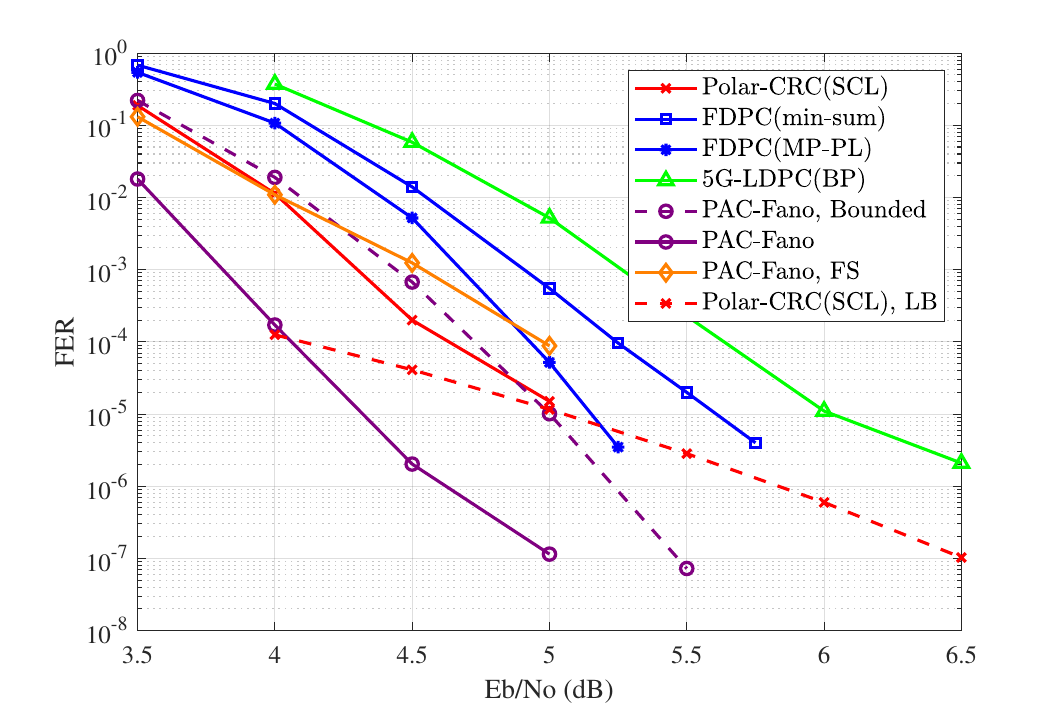}
	\caption{FER performance comparison of PAC, polar, LDPC, and FDPC codes with a block length of 1024 and a code rate of 0.878.} 
	\label{fig: N1024_HighRate_Guessing_FER}
\end{figure}

\begin{figure}[htbp] 
\centering
	\includegraphics[width = \columnwidth]{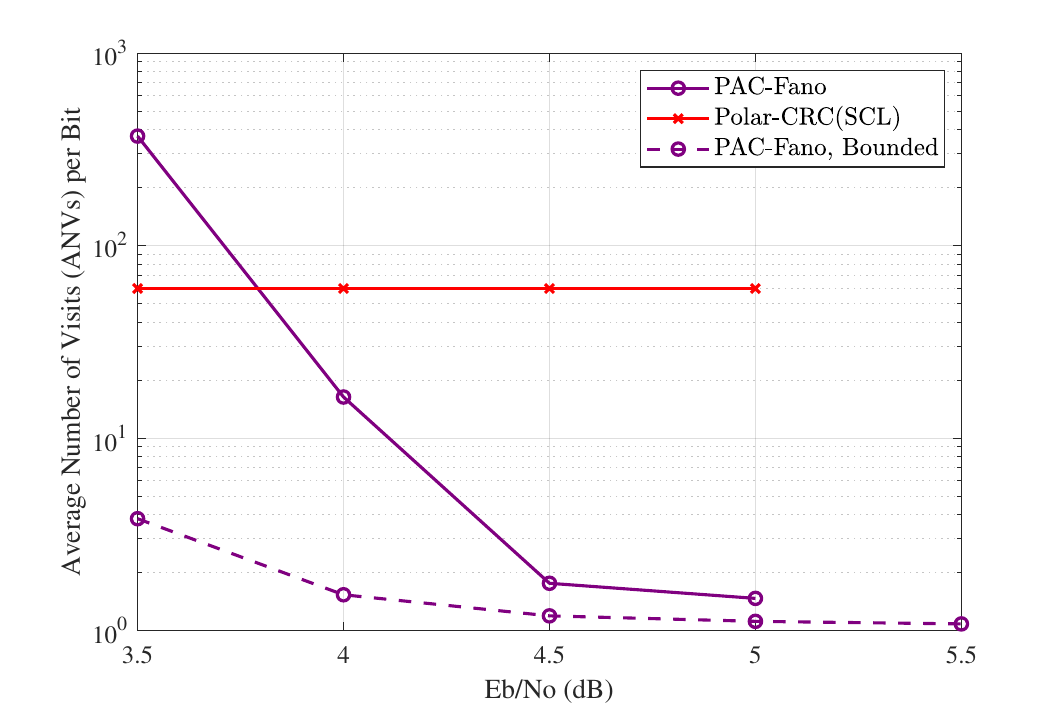}
	\caption{Complexity performance comparison of PAC and CRC-aided polar codes with a block length of 1024 and a code rate of
0.878.} 
	\label{fig: N1024_HighRate_Guessing_ANV}
\end{figure}

The performance of the designed PAC\,$(1024, 899)$ code is illustrated in Fig. \ref{fig: N1024_HighRate_Guessing_FER}, where it is compared to the corresponding 5G-polar code with CRC-aided list decoding (list size of 32) and the 5G-LDPC code with BP decoding. 
Additionally, the figure includes the FER performance of the fair-density parity-check (FDPC) code with min-sum and MP-PL decoding \cite{mahdavifar2024high, moradi2025highFDPC}.
Our designed PAC code demonstrates a coding gain of over 0.75 dB at a FER of $10^{-5}$ when compared to other state-of-the-art codes. 
In this figure, we also include the performance of our designed PAC code when using a Fano decoder with a bounded amount of search. The maximum number of node visits on the decoding tree is set to $10\times N = 10240$ per codeword.
We use $\Delta = 6.5$ in this case. 
Although our theoretical results suggest using $\Delta = 2$ when the amount of search is not bounded, our numerical investigation suggests using a higher value of the threshold spacing $\Delta$ when the maximum number of visits is decreased.  
For an SCL decoding of $(1024, 899)$ code, the maximum number of visits is $2+4+8+16+32+(899-5)\times64+(1024-899)\times32 = 61278$.
A fair comparison of the maximum and average latency and complexity performance comparision of SCL decoding and Fano decoding requires further investigation as a Fano decoding sequentially decodes the bits and explores the decoding tree back and forth which adds to the latency of the decoding.
In Fig. \ref{fig: N1024_HighRate_Guessing_FER}, we also show the performance of Fano decoding of PAC codes when the freezing bits correspond to the upper rows (PAC-Fano, FS), as explained previously.
Furthermore, Fig. \ref{fig: N1024_HighRate_Guessing_ANV} depicts the average number of visits (ANV) per decoded bit for the Fano decoding of our design when exploring the decoding tree. 
As shown, at high \( E_b/N_0 \) values, the number of visits at each level of the decoding tree approaches 1, highlighting the efficiency and practicality of the proposed design.
In this figure, we also include the performance of our designed PAC code when using a Fano decoder with a bounded search. The maximum number of visits for the bounded-search Fano decoding is 10 per bit on average.
For comparison, the figure also shows the average number of visits for SCL decoding of the CRC-aided code.

\begin{figure}[htbp] 
\centering
	\includegraphics[width = \columnwidth]{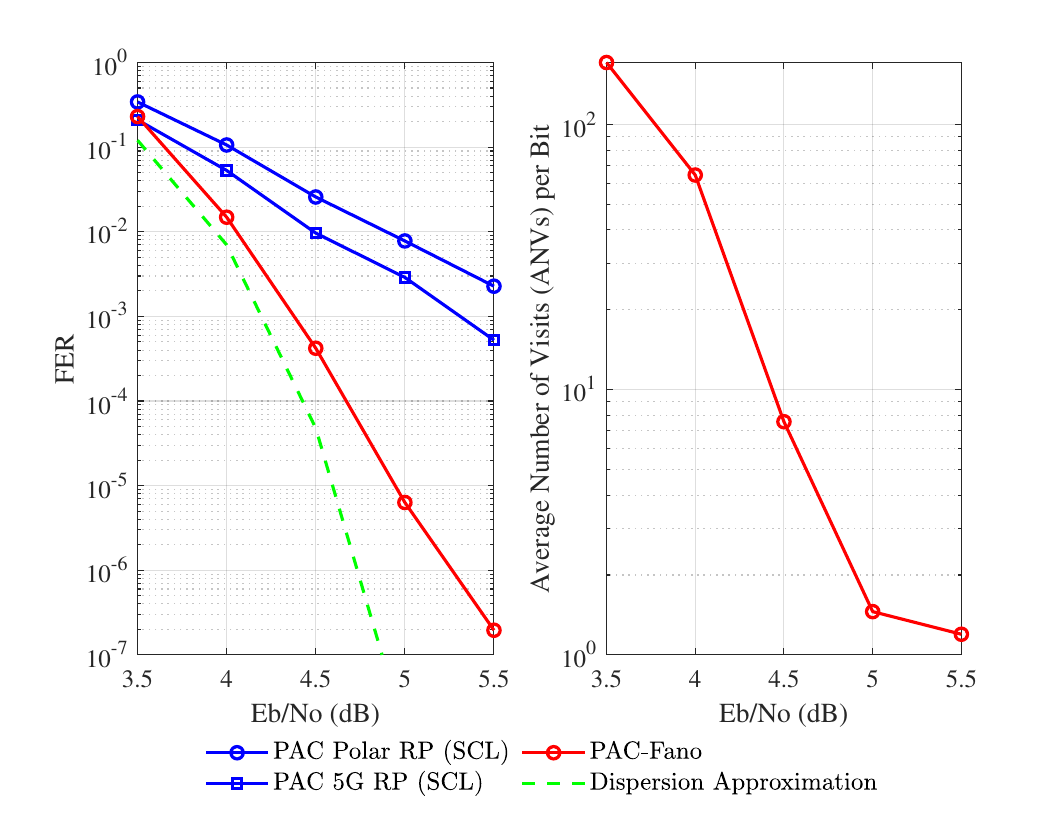}
	\caption{Performance comparison of the PAC\,$(512, 460)$ codes with Fano and SCL decoding algorithms with different rate profiles.} 
	\label{fig: N512_K460_HighRate_Guessing}
\end{figure}

Fig. \ref{fig: N512_K460_HighRate_Guessing} also plots the performance of our rate profile construction of a PAC\,$(512, 460)$ of a rate almost equal to 0.9 code with Fano decoding algorithm. 
We begin with an RM$(512,466)$ code, which has a minimum distance of 8.
We assume a polarization step of \( k = 6 \), resulting in 64 chunks, each of length 8. 
At a target \( E_b/N_0 = 4.5~\text{dB} \), we must freeze at least 5 of the data bits from the RM$(512, 466)$ code to satisfy the constraint imposed by the polarization cutoff rate, resulting in a $(512, 461)$ code. 
We then freeze 1 data bit, corresponding to the last rows of the generator matrix with weight 8 of obtained (512,461) code, to construct a PAC$(512, 460)$ code.

The performance of the designed PAC$(512, 460)$ code is illustrated in Fig. \ref{fig: N512_K460_HighRate_Guessing}, where it is compared to the corresponding PAC codes with 5G rate profile and polar rate profile constructed at $4.5~$dB with a list decoding (list size of 32). 
Additionally, the figure includes the plot of dispersion approximation.
Our designed PAC code demonstrates a coding gain of about 1 dB at a FER of $10^{-3}$ when compared to other codes. 
Furthermore, Fig. \ref{fig: N512_K460_HighRate_Guessing} depicts the ANV per decoded bit for the Fano decoding of our design when exploring the decoding tree.

Note that at the target \( E_b/N_0 = 4.5~\text{dB} \), the channel cutoff rate is \( 0.8931 \), indicating that we can use at most about 457 data bits to have low-complexity sequential decoding for convolutional codes. However, by the 6th step of polarization, the polarized cutoff rates allow up to 479 data bits for low-complexity sequential decoding for PAC codes, providing a gain in code rate by boosting the cutoff rate. Our proposed algorithm selects 460 data-bit locations to achieve good FER performance, utilizing an RM code rate profile.

\begin{figure}[htbp] 
\centering
	\includegraphics[width = \columnwidth]{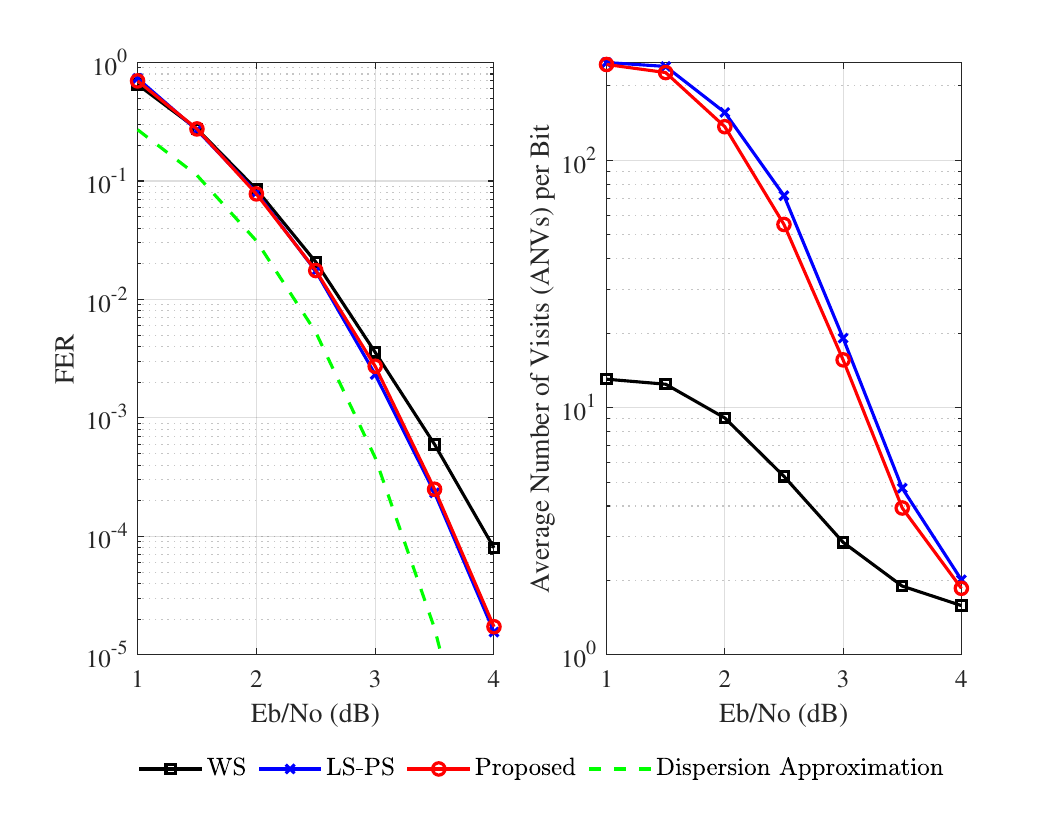}
	\caption{Performance comparison of the PAC\,$(128, 85)$ codes with Fano decoding algorithms with different rate profiles.} 
	\label{fig: N128_K80_HighRate_Guessing}
\end{figure}

Fig.~\ref{fig: N128_K80_HighRate_Guessing} also plots the performance of our rate profile construction for a PAC$(128, 85)$ code using the Fano decoding algorithm. 
We begin with an RM$(128, 99)$ code, which has a minimum distance of 8. We assume a polarization step of \( k = 4 \), resulting in 16 chunks, each of length 8. 
At a target \( E_b/N_0 = 3~\text{dB} \), we must freeze at least 4 data bits from the RM$(128, 99)$ code to satisfy the constraint imposed by the polarization cutoff rate, resulting in a $(128, 95)$ code. 
We then freeze 10 additional data bits, corresponding to the last rows of the generator matrix with weight 8 of the resulting $(128, 95)$ code, to construct a PAC$(128, 85)$ code.
In this figure, we also plot the performance of the Fano decoding of the weighted sum-based PAC code construction proposed in~\cite{liu2022weighted}, as well as the performance of the PAC code constructed using list-search and path-splitting (LS-PS) critical sets from~\cite{jiang2023construction}.

%%%%%%%%%%%%%%%%%%%%%%%%%%%%%%%%%%%%%%%%%%%%%%%%%%%%%%%%%%%%%%%%%%%%%%%%%%%%%%%%%%%%%%%%%%%%%%%%%%%%%%%
\section{Conclusion}\label{sec: Conclusion}\label{sec: conclusion}
In sequential decoding of convolutional codes, when the code rate falls below the channel cutoff rate, the computation complexity follows a Pareto distribution. 
In this paper, we explore the computational complexity of sequential decoding when applied to PAC codes, and show that if the polarized rate of the code is less than the polarized cutoff rate, the computation complexity for the corresponding portion of the decoding tree exhibits a Pareto distribution. 
Leveraging this result, we propose a rate-profile construction method that can achieve superior error-correction performance compared to state-of-the-art codes.

%%%%%%%%%%%%%%%%%%%%%%%%%%%%%%%%%%%%%%%%%%%%%%%%%%%%%%%%%%%%%%%%%%%%%%%%%%%%%%%%%%%%%%%%%%%%%%%%%%%%%%%
% \section*{Acknowledgment}

\ifCLASSOPTIONcaptionsoff
  \newpage
\fi

\bibliographystyle{IEEEtran}
\bibliography{bibliography}

\begin{thebibliography}{10}
\providecommand{\url}[1]{#1}
\csname url@samestyle\endcsname
\providecommand{\newblock}{\relax}
\providecommand{\bibinfo}[2]{#2}
\providecommand{\BIBentrySTDinterwordspacing}{\spaceskip=0pt\relax}
\providecommand{\BIBentryALTinterwordstretchfactor}{4}
\providecommand{\BIBentryALTinterwordspacing}{\spaceskip=\fontdimen2\font plus
\BIBentryALTinterwordstretchfactor\fontdimen3\font minus \fontdimen4\font\relax}
\providecommand{\BIBforeignlanguage}[2]{{%
\expandafter\ifx\csname l@#1\endcsname\relax
\typeout{** WARNING: IEEEtran.bst: No hyphenation pattern has been}%
\typeout{** loaded for the language `#1'. Using the pattern for}%
\typeout{** the default language instead.}%
\else
\language=\csname l@#1\endcsname
\fi
#2}}
\providecommand{\BIBdecl}{\relax}
\BIBdecl

\bibitem{arikan2019sequential}
\BIBentryALTinterwordspacing
E.~Arıkan, ``From sequential decoding to channel polarization and back again,'' 2019. [Online]. Available: \url{https://arxiv.org/abs/1908.09594}
\BIBentrySTDinterwordspacing

\bibitem{moradi2020PAC}
\BIBentryALTinterwordspacing
M.~Moradi, A.~Mozammel, K.~Qin, and E.~Ar{\i}kan, ``Performance and complexity of sequential decoding of {PAC} codes,'' 2020. [Online]. Available: \url{https://arxiv.org/abs/2012.04990}
\BIBentrySTDinterwordspacing

\bibitem{arikan2009channel}
E.~Ar{\i}kan, ``Channel polarization: {A} method for constructing capacity-achieving codes for symmetric binary-input memoryless channels,'' \emph{IEEE Transactions on Information Theory}, vol.~55, no.~7, pp. 3051--3073, 2009.

\bibitem{wozencraft1957sequential}
J.~M. Wozencraft, ``Sequential decoding for reliable communication,'' Research Laboratory of Electronics, MIT, Cambridge, Tech. Rep. 325, 1957.

\bibitem{fano1963heuristic}
R.~Fano, ``A heuristic discussion of probabilistic decoding,'' \emph{IEEE Transactions on Information Theory}, vol.~9, no.~2, pp. 64--74, 1963.

\bibitem{massey1972variable}
J.~Massey, ``Variable-length codes and the {F}ano metric,'' \emph{IEEE Transactions on Information Theory}, vol.~18, no.~1, pp. 196--198, 1972.

\bibitem{moradi2021sequential}
M.~Moradi, ``On sequential decoding metric function of polarization-adjusted convolutional ({PAC}) codes,'' \emph{IEEE Transactions on Communications}, vol.~69, no.~12, pp. 7913--7922, 2021.

\bibitem{moradi2024fast}
\BIBentryALTinterwordspacing
M.~Moradi and H.~Mahdavifar, ``On fast {SC}-based polar decoders: Metric polarization and a pruning technique,'' 2024. [Online]. Available: \url{https://arxiv.org/abs/2408.03840}
\BIBentrySTDinterwordspacing

\bibitem{niu2012stack}
K.~Niu and K.~Chen, ``Stack decoding of polar codes,'' \emph{Electronics letters}, vol.~48, no.~12, pp. 695--697, 2012.

\bibitem{miloslavskaya2014sequential}
V.~Miloslavskaya and P.~Trifonov, ``Sequential decoding of polar codes,'' \emph{IEEE Communications Letters}, vol.~18, no.~7, pp. 1127--1130, 2014.

\bibitem{trifonov2018score}
P.~Trifonov, ``A score function for sequential decoding of polar codes,'' in \emph{2018 IEEE International Symposium on Information Theory (ISIT)}, 2018, pp. 1470--1474.

\bibitem{yao2021list}
H.~Yao, A.~Fazeli, and A.~Vardy, ``List decoding of {Ar{\i}kan’s PAC} codes,'' \emph{Entropy}, vol.~23, no.~7, p. 841, 2021.

\bibitem{rowshan2021polarization}
M.~Rowshan, A.~Burg, and E.~Viterbo, ``Polarization-adjusted convolutional {(PAC)} codes: Sequential decoding vs list decoding,'' \emph{IEEE Transactions on Vehicular Technology}, vol.~70, no.~2, pp. 1434--1447, 2021.

\bibitem{seyedmasoumian2022approximate}
S.~Seyedmasoumian and T.~M. Duman, ``Approximate weight distribution of polarization-adjusted convolutional {(PAC)} codes,'' in \emph{2022 IEEE International Symposium on Information Theory (ISIT)}.\hskip 1em plus 0.5em minus 0.4em\relax IEEE, 2022, pp. 2577--2582.

\bibitem{rowshan2023minimum}
M.~Rowshan and J.~Yuan, ``On the minimum weight codewords of {PAC} codes: The impact of pre-transformation,'' \emph{IEEE Journal on Selected Areas in Information Theory}, 2023.

\bibitem{dumer2006soft}
I.~Dumer and K.~Shabunov, ``Soft-decision decoding of {R}eed-{M}uller codes: recursive lists,'' \emph{IEEE Transactions on Information Theory}, vol.~52, no.~3, pp. 1260--1266, 2006.

\bibitem{li2014rm}
B.~Li, H.~Shen, and D.~Tse, ``A {RM}-polar codes,'' \emph{arXiv preprint arXiv:1407.5483}, 2014.

\bibitem{moradi2023application}
M.~Moradi, ``Application of guessing to sequential decoding of polarization-adjusted convolutional ({PAC}) codes,'' \emph{IEEE Transactions on Communications}, vol.~71, no.~8, pp. 4425--4436, 2023.

\bibitem{miloslavskaya2023design}
V.~Miloslavskaya, Y.~Li, and B.~Vucetic, ``Design of compactly specified polar codes with dynamic frozen bits based on reinforcement learning,'' \emph{IEEE Transactions on Communications}, vol.~72, no.~3, pp. 1257--1272, 2024.

\bibitem{trifonov2013polar}
P.~Trifonov and V.~Miloslavskaya, ``Polar codes with dynamic frozen symbols and their decoding by directed search,'' in \emph{2013 IEEE Information Theory Workshop (ITW)}, 2013, pp. 1--5.

\bibitem{wang2023road}
C.-X. Wang, X.~You, X.~Gao, X.~Zhu, Z.~Li, C.~Zhang, H.~Wang, Y.~Huang, Y.~Chen, H.~Haas \emph{et~al.}, ``On the road to {6G}: Visions, requirements, key technologies, and testbeds,'' \emph{IEEE Communications Surveys \& Tutorials}, vol.~25, no.~2, pp. 905--974, 2023.

\bibitem{savage1965computation}
J.~E. Savage, ``The computation problem with sequential decoding.'' Ph.D. dissertation, ept. of Elec. Engrg., Massachusetts Institute of Technology, Cambridge, February 1965.

\bibitem{jacobs}
I.~M. Jacobs and J.~Wozencraft, \emph{Principles of communication engineering.}\hskip 1em plus 0.5em minus 0.4em\relax New York: John Wiley and Sons, 1965.

\bibitem{gallager1968information}
R.~G. Gallager, \emph{Information theory and reliable communication}.\hskip 1em plus 0.5em minus 0.4em\relax New York: Wiley, 1968, vol.~2.

\bibitem{arikan1996inequality}
E.~Ar{\i}kan, ``An inequality on guessing and its application to sequential decoding,'' \emph{IEEE Transactions on Information Theory}, vol.~42, no.~1, pp. 99--105, 1996.

\bibitem{mozammel2021hardware}
A.~Mozammel, ``Hardware implementation of {F}ano decoder for polarization-adjusted convolutional ({PAC}) codes,'' \emph{IEEE Transactions on Circuits and Systems II: Express Briefs}, vol.~69, no.~3, pp. 1632--1636, 2022.

\bibitem{moradi2023tree}
M.~Moradi and A.~Mozammel, ``A tree pruning technique for decoding complexity reduction of polar codes and {PAC} codes,'' \emph{IEEE Transactions on Communications}, vol.~71, no.~5, pp. 2576--2586, 2023.

\bibitem{gallager2013stochastic}
R.~G. Gallager, \emph{Stochastic {Processes}: {Theory} for {Applications}}.\hskip 1em plus 0.5em minus 0.4em\relax Cambridge University Press, 2013.

\bibitem{cocskun2022information}
M.~C. Coşkun and H.~D. Pfıster, ``An information-theoretic perspective on successive cancellation list decoding and polar code design,'' \emph{IEEE Transactions on Information Theory}, vol.~68, no.~9, pp. 5779--5791, 2022.

\bibitem{mahdavifar2024high}
\BIBentryALTinterwordspacing
H.~Mahdavifar, ``High-rate fair-density parity-check codes,'' 2024. [Online]. Available: \url{https://arxiv.org/abs/2402.06814}
\BIBentrySTDinterwordspacing

\bibitem{moradi2025highFDPC}
\BIBentryALTinterwordspacing
M.~Moradi, S.~Rabeti, and H.~Mahdavifar, ``On the high-rate {FDPC} codes: Construction, encoding, and a generalization,'' 2025. [Online]. Available: \url{https://arxiv.org/abs/2506.11345}
\BIBentrySTDinterwordspacing

\bibitem{liu2022weighted}
W.~Liu, L.~Chen, and X.~Liu, ``A weighted sum based construction of pac codes,'' \emph{IEEE Communications Letters}, vol.~27, no.~1, pp. 28--31, 2023.

\bibitem{jiang2023construction}
S.~Jiang, J.~Wang, C.~Xia, and X.~Li, ``Construction of pac codes with list-search and path-splitting critical sets,'' \emph{arXiv preprint arXiv:2304.11554}, 2023.

\end{thebibliography}

\end{document}